\documentclass[%
reprint,
superscriptaddress,
nofootinbib,
amsmath,amssymb,
twocolumn,
aps,
prl,
floatfix]{revtex4-2}

\usepackage{xcolor}
\usepackage{subfigure}
\usepackage{graphicx}
\usepackage{dcolumn}
\usepackage{bm}
\usepackage[pdftex]{hyperref} 
\usepackage[mathlines]{lineno}
\usepackage{enumerate}%

\usepackage{svg}

\usepackage{adjustbox}
\usepackage{multirow}
\usepackage{lineno}

\newcommand{\MA}{M_{A'}}

\newcommand{\ee}{e^+e^-}

\newcommand{\RM}{M_{\chi}/M_{A'}}

\begin{document}


\widetext

\title{First search for dark-trident processes using the MicroBooNE detector}%

\newcommand{\ANL}{Argonne National Laboratory (ANL), Lemont, IL, 60439, USA}
\newcommand{\Bern}{Universit{\"a}t Bern, Bern CH-3012, Switzerland}
\newcommand{\BNL}{Brookhaven National Laboratory (BNL), Upton, NY, 11973, USA}
\newcommand{\UCSB}{University of California, Santa Barbara, CA, 93106, USA}
\newcommand{\Cambridge}{University of Cambridge, Cambridge CB3 0HE, United Kingdom}
\newcommand{\CIEMAT}{Centro de Investigaciones Energ\'{e}ticas, Medioambientales y Tecnol\'{o}gicas (CIEMAT), Madrid E-28040, Spain}
\newcommand{\Chicago}{University of Chicago, Chicago, IL, 60637, USA}
\newcommand{\Cincinnati}{University of Cincinnati, Cincinnati, OH, 45221, USA}
\newcommand{\CSU}{Colorado State University, Fort Collins, CO, 80523, USA}
\newcommand{\Columbia}{Columbia University, New York, NY, 10027, USA}
\newcommand{\Edinburgh}{University of Edinburgh, Edinburgh EH9 3FD, United Kingdom}
\newcommand{\FNAL}{Fermi National Accelerator Laboratory (FNAL), Batavia, IL 60510, USA}
\newcommand{\Granada}{Universidad de Granada, Granada E-18071, Spain}
\newcommand{\Harvard}{Harvard University, Cambridge, MA 02138, USA}
\newcommand{\IIT}{Illinois Institute of Technology (IIT), Chicago, IL 60616, USA}
\newcommand{\Indiana}{Indiana University, Bloomington, IN 47405, USA}
\newcommand{\KSU}{Kansas State University (KSU), Manhattan, KS, 66506, USA}
\newcommand{\Lancaster}{Lancaster University, Lancaster LA1 4YW, United Kingdom}
\newcommand{\LANL}{Los Alamos National Laboratory (LANL), Los Alamos, NM, 87545, USA}
\newcommand{\Louisiana}{Louisiana State University, Baton Rouge, LA, 70803, USA}
\newcommand{\Manchester}{The University of Manchester, Manchester M13 9PL, United Kingdom}
\newcommand{\MIT}{Massachusetts Institute of Technology (MIT), Cambridge, MA, 02139, USA}
\newcommand{\Michigan}{University of Michigan, Ann Arbor, MI, 48109, USA}
\newcommand{\MSU}{Michigan State University, East Lansing, MI 48824, USA}
\newcommand{\Minnesota}{University of Minnesota, Minneapolis, MN, 55455, USA}
\newcommand{\Nankai}{Nankai University, Nankai District, Tianjin 300071, China}
\newcommand{\NMSU}{New Mexico State University (NMSU), Las Cruces, NM, 88003, USA}
\newcommand{\Oxford}{University of Oxford, Oxford OX1 3RH, United Kingdom}
\newcommand{\Pitt}{University of Pittsburgh, Pittsburgh, PA, 15260, USA}
\newcommand{\Rutgers}{Rutgers University, Piscataway, NJ, 08854, USA}
\newcommand{\SLAC}{SLAC National Accelerator Laboratory, Menlo Park, CA, 94025, USA}
\newcommand{\SDSMT}{South Dakota School of Mines and Technology (SDSMT), Rapid City, SD, 57701, USA}
\newcommand{\Maine}{University of Southern Maine, Portland, ME, 04104, USA}
\newcommand{\Syracuse}{Syracuse University, Syracuse, NY, 13244, USA}
\newcommand{\TelAviv}{Tel Aviv University, Tel Aviv, Israel, 69978}
\newcommand{\Tennessee}{University of Tennessee, Knoxville, TN, 37996, USA}
\newcommand{\UTA}{University of Texas, Arlington, TX, 76019, USA}
\newcommand{\Tufts}{Tufts University, Medford, MA, 02155, USA}
\newcommand{\UCL}{University College London, London WC1E 6BT, United Kingdom}
\newcommand{\VTech}{Center for Neutrino Physics, Virginia Tech, Blacksburg, VA, 24061, USA}
\newcommand{\Warwick}{University of Warwick, Coventry CV4 7AL, United Kingdom}
\newcommand{\Yale}{Wright Laboratory, Department of Physics, Yale University, New Haven, CT, 06520, USA}

\affiliation{\ANL}
\affiliation{\Bern}
\affiliation{\BNL}
\affiliation{\UCSB}
\affiliation{\Cambridge}
\affiliation{\CIEMAT}
\affiliation{\Chicago}
\affiliation{\Cincinnati}
\affiliation{\CSU}
\affiliation{\Columbia}
\affiliation{\Edinburgh}
\affiliation{\FNAL}
\affiliation{\Granada}
\affiliation{\Harvard}
\affiliation{\IIT}
\affiliation{\Indiana}
\affiliation{\KSU}
\affiliation{\Lancaster}
\affiliation{\LANL}
\affiliation{\Louisiana}
\affiliation{\Manchester}
\affiliation{\MIT}
\affiliation{\Michigan}
\affiliation{\MSU}
\affiliation{\Minnesota}
\affiliation{\Nankai}
\affiliation{\NMSU}
\affiliation{\Oxford}
\affiliation{\Pitt}
\affiliation{\Rutgers}
\affiliation{\SLAC}
\affiliation{\SDSMT}
\affiliation{\Maine}
\affiliation{\Syracuse}
\affiliation{\TelAviv}
\affiliation{\Tennessee}
\affiliation{\UTA}
\affiliation{\Tufts}
\affiliation{\UCL}
\affiliation{\VTech}
\affiliation{\Warwick}
\affiliation{\Yale}

\author{P.~Abratenko} \affiliation{\Tufts}
\author{O.~Alterkait} \affiliation{\Tufts}
\author{D.~Andrade~Aldana} \affiliation{\IIT}
\author{L.~Arellano} \affiliation{\Manchester}
\author{J.~Asaadi} \affiliation{\UTA}
\author{A.~Ashkenazi}\affiliation{\TelAviv}
\author{S.~Balasubramanian}\affiliation{\FNAL}
\author{B.~Baller} \affiliation{\FNAL}
\author{G.~Barr} \affiliation{\Oxford}
\author{D.~Barrow} \affiliation{\Oxford}
% ...

\collaboration{The MicroBooNE Collaboration}
\thanks{microboone\_info@fnal.gov}\noaffiliation


\begin{abstract}
We present a first search for dark-trident scattering in a neutrino beam using a data set
corresponding to $7.2 \times 10^{20}$ protons on target taken with the
MicroBooNE detector at Fermilab. 
Proton interactions in the neutrino target at the Main Injector produce $\pi^0$ and $\eta$ mesons, which could
decay into dark-matter (DM) particles mediated via a dark photon $A^\prime$. 
A convolutional neural network is trained to identify
interactions of the DM particles in the liquid-argon time projection chamber (LArTPC) exploiting its image-like
reconstruction capability. 
In the absence of a DM signal, we provide limits at the $90\%$ confidence level on the squared kinematic mixing parameter $\varepsilon^2$ as a function of the
dark-photon mass in the range $10\le M_{A^\prime}\le 400$~MeV. The limits cover
 previously unconstrained parameter space for
the production of fermion or scalar 
DM particles $\chi$ for two benchmark models with
mass ratios $M_{\chi}/M_{A^\prime}=0.6$ and $2$ and
for dark fine-structure constants $0.1\le\alpha_D\le 1$. 
\end{abstract}
                        
\maketitle
A wealth of astronomical data at different scales 
provide evidence for the existence
of dark matter (DM): the motion of galaxies and the stars within them, gravitational lensing, the cosmic microwave background, and the large-scale structure of the universe~\cite{Planck:2018vyg}.
The nature of dark matter, however, remains elusive. Non-baryonic particles predicted by dark-sector models are candidates for dark matter~\cite{Arbey:2021gdg}. The search for their production at accelerators is a focus of the high-energy hadron collider program at the LHC~\cite{Rizzo:2022qan} and of fixed-target
experiments exposed to high-intensity beams~\cite{Antel:2023hkf}. 

The dark-trident process has been proposed as a new way to search for low-mass dark-matter particles
in neutrino beams~\cite{deGouvea:2018cfv}. In this letter, we report a first search for such dark tridents 
with the MicroBooNE liquid-argon time projection chamber (LArTPC)~\cite{Acciarri:2016smi}. 
In the future, similar searches can be performed 
with the DUNE near detector~\cite{DUNE:2020fgq} and the detectors of the Fermilab short-baseline program~\cite{Machado:2019oxb}.

\begin{figure}[htbp]
    \centering
    \includegraphics[width=0.35\textwidth]{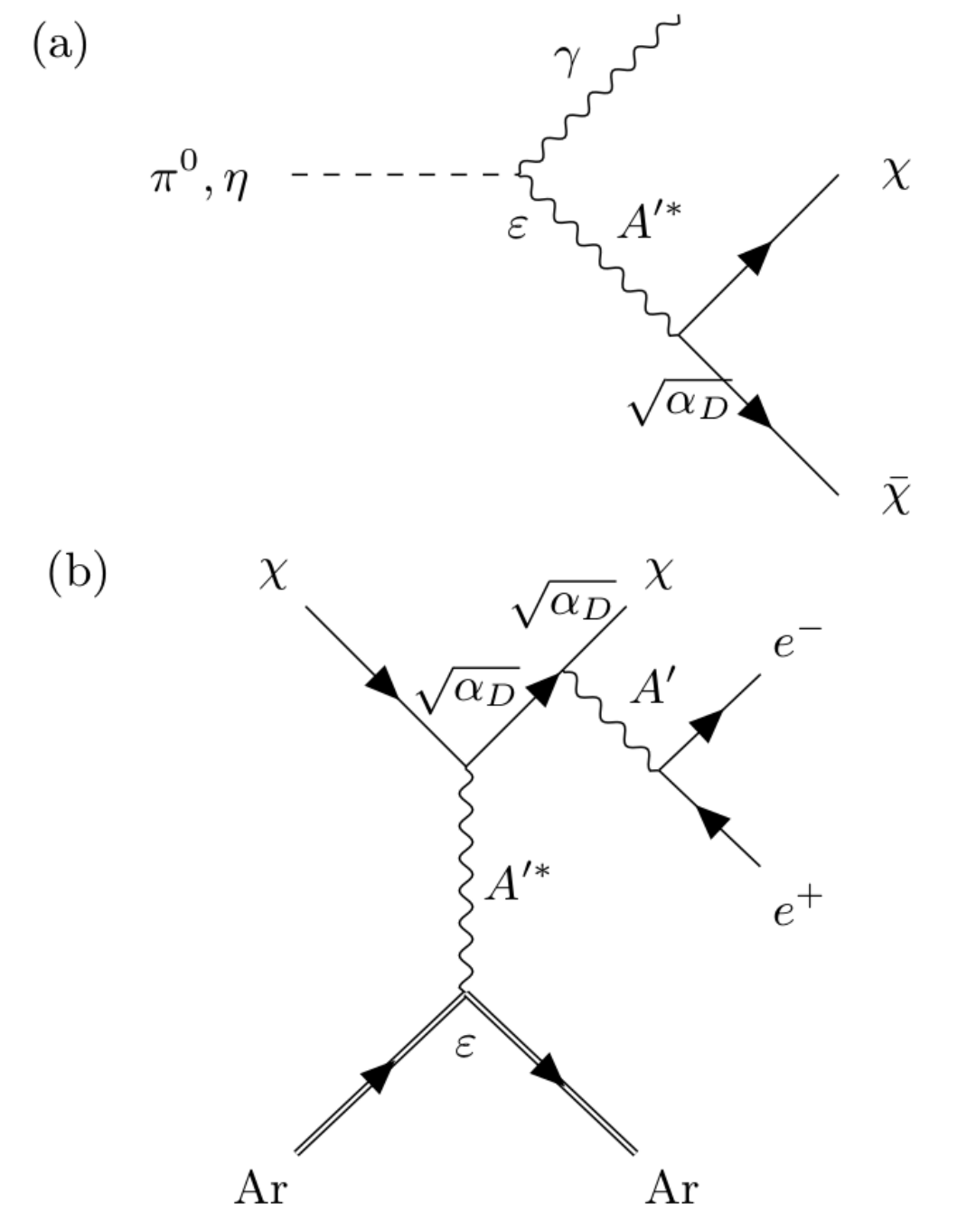}
    \caption{(a) A pair of DM particles, $\chi\bar{\chi}$, is produced in a 
    $\pi^{0}$ or $\eta^0$ decay; (b) in the dark-trident process, $\chi$ (or $\bar{\chi}$) scatters off an argon nucleus to produce a dark photon $A'$ decaying into an $\ee$ pair with a branching ratio of $1$. The rate depends on the
    kinematic mixing parameter $\varepsilon$ and the dark fine-structure constant~$\alpha_{D}$.
    }
\label{fig:diag}
\vskip -5mm
\end{figure}

A pair of DM particles, $\chi\bar{\chi}$, is produced through the decay of neutral $\pi^{0}$ or $\eta$ mesons, which was created by the interactions of the protons and by secondary interactions in the neutrino target (Fig.~\ref{fig:diag}a). The decays $\pi^{0},\eta\rightarrow \gamma \chi \bar{\chi}$ are mediated by a virtual, off-shell dark photon $A'^*$. The masses of the dark photon, $\MA$, and of the dark fermion (or scalar), $M_{\chi}$, are parameters of the model. The energies of the DM particles $\chi$ are typically in the
range $0.1$--$3$~GeV for the mass range $10<M_{\chi}<400$~MeV.

\begin{figure}[htbp]
    \centering
       \includegraphics[width=0.35\textwidth]{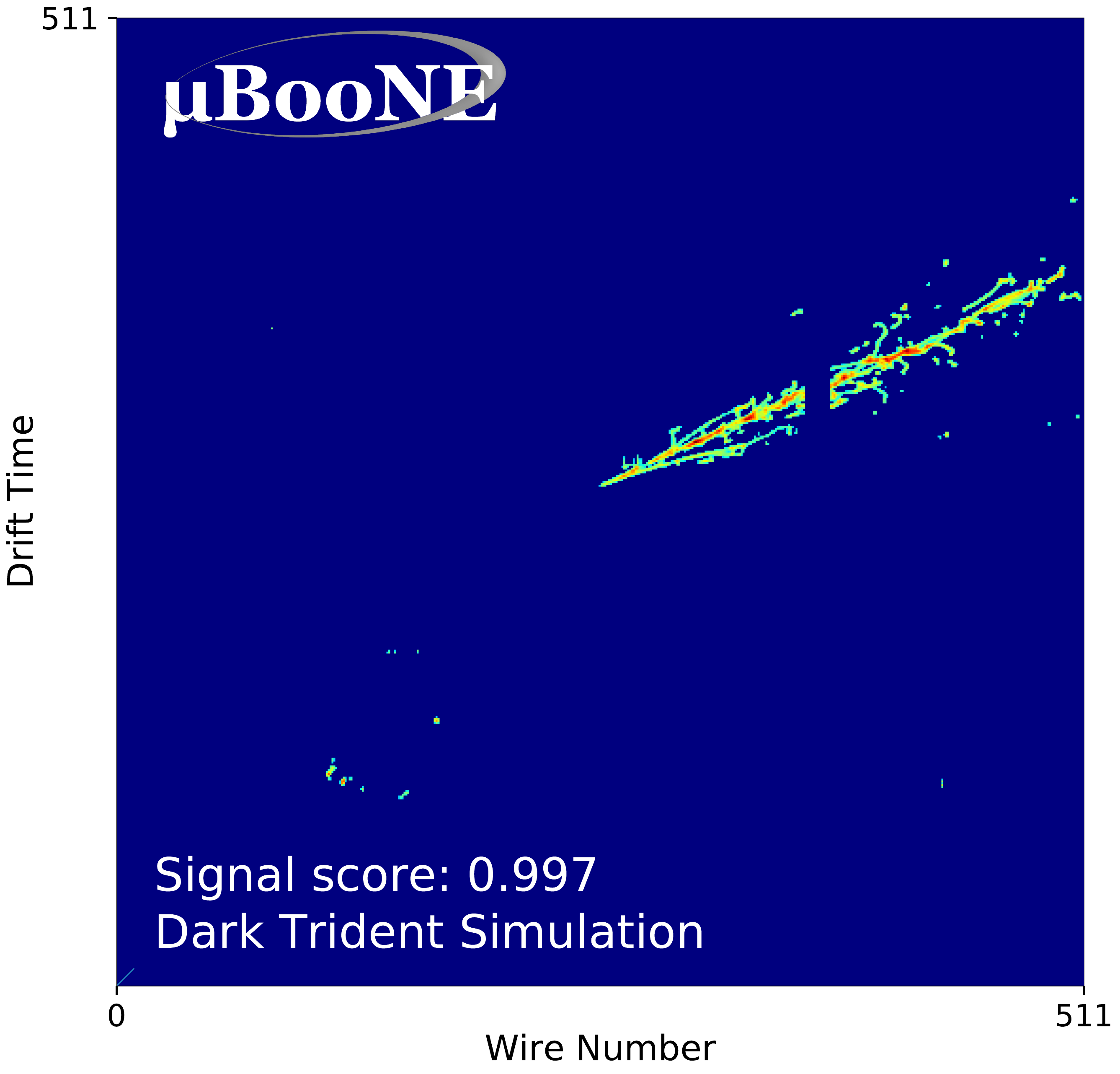} 
       \includegraphics[width=0.35\textwidth]{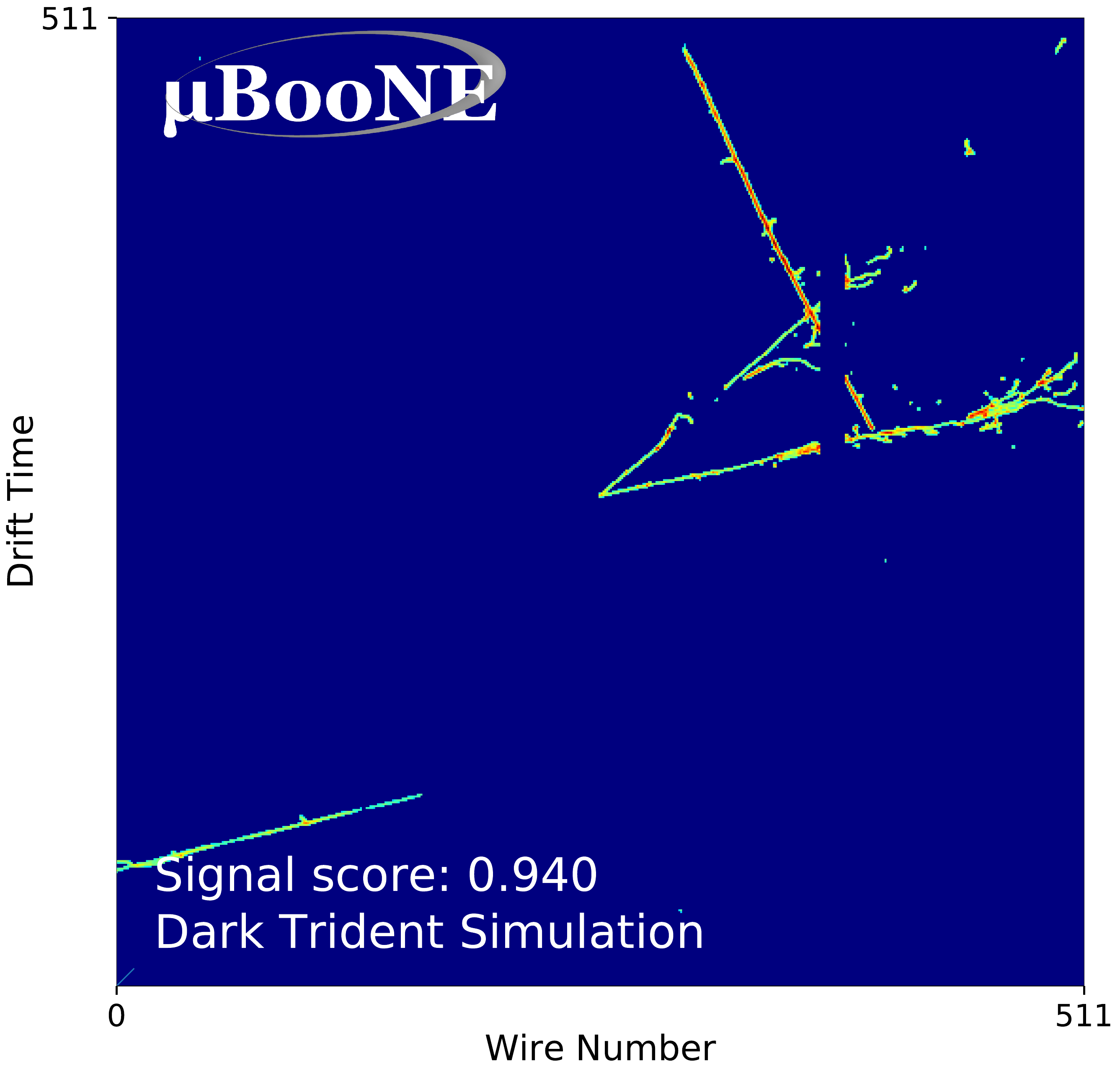} 
           \caption{\label{fig:sim_displays} Simulated dark-trident interactions in the MicroBooNE detector assuming dark photon masses of $M_{A'}=50$~MeV (top) and $300$~MeV (bottom). The horizontal gaps are due to unresponsive wires.}
\end{figure}

The DM particle $\chi$ (or $\bar{\chi}$) then travels uninterrupted to the MicroBooNE detector where it could scatter off argon nuclei through the trident process $\chi + \textrm{Ar} \rightarrow \chi + \textrm{Ar} + A'$~(see Fig.~\ref{fig:diag}b). The dark photon $A'$ promptly decays inside the detector into an $\ee$ pair. The energies and opening angles of the $\ee$ pairs depend on the ratio $\RM$ (see Fig.~\ref{fig:sim_displays}). The $\chi$ production rate depends on the kinematic mixing parameter $\varepsilon$ and a dark fine-structure constant $\alpha_{D}$, which is defined in terms of the 
dark-photon gauge coupling $g_D$ as $\alpha_D= g_D^{2}/(4\pi)$.
We consider the mass ratios $\RM=0.6$ and $2$ in this search as proposed in Ref.~\cite{deGouvea:2018cfv}. Since $\RM>0.5$,
the dark photons need to be off-shell to decay into $\chi\bar{\chi}$ pairs, and, when on-shell, they will exclusively decay to $\ee$. The signal 
rate therefore scales with $\varepsilon^4\alpha_{D}^3$. 

In this letter, we study dark photon masses in the range $10<M_{A'}<400$~MeV, since the parameter space below $M_{A'}=10$~MeV is constrained by beam dump searches~\cite{Bjorken:1988as,Riordan:1987aw,Bross:1989mp}. The decay $\eta \rightarrow \chi \bar\chi \gamma $ is kinematically forbidden for $M_{A'} > 457$ MeV (assuming  $M_\chi/M_{A'} =0.6$). Other recent experimental searches cover the mass range where $A'$ decays invisibly~\cite{PhysRevD.98.112004, PhysRevLett.130.051803, NA64:2023wbi}. 

The MicroBooNE's LArTPC has an instrumented volume of $85$~tonnes of liquid argon inside a cryostat. 
Ionization charge drifts across an electric field of $273$~V/cm and is read out by one charge collection and two induction planes forming the anode. 
The LArTPC was simultaneously exposed to the on-axis booster neutrino beam (BNB)~\cite{AguilarArevalo:2008yp} and
the off-axis beam of neutrinos from the main injector (NuMI)~\cite{ADAMSON2016279}. Only NuMI data are used in this 
search, as the higher average energy of the NuMI beam gives access to higher values of $\MA$.  
 
The NuMI data used for this analysis~\cite{LuisMoraLepin} correspond to $7.2 \times 10^{20}$ protons on target (POT), which were taken in two operating modes -- forward horn current 
(FHC) with $2.2 \times 10^{20}$~POT (Run~1 from October 2015 to November 2017) and reverse horn current (RHC) with $5.0 \times 10^{20}$~POT (Run~3 from November 2017 to July 2018). 
This data set has previously been used to search for heavy neutral leptons~\cite{MicroBooNE:2023icy,MicroBooNE:2022ctm} and Higgs portal scalars~\cite{MicroBooNE:2022ctm,MicroBooNE:2021usw},
and to measure neutrino cross sections~\cite{MicroBooNE:2021gfj,MicroBooNE:2021ppm}.

We simulate the dark-trident process with a dedicated generator in three steps:
the neutral meson flux in the beamline, the decay of the 
neutral mesons, and the scattering of the DM particles on argon.
First, the kinematics of the $\pi^{0}$ and $\eta$ mesons for both beam configurations, FHC and RHC, are obtained using the \texttt{g4NuMI} simulation~\cite{Aliaga:2016oaz}, which is based on 
a full \texttt{GEANT4} description of the beamline geometry. The \texttt{g4NuMI} simulation predicts $\approx 32$~$\pi^{0}$ and $\approx 2.5$~$\eta$ mesons per POT, compared to
$4.5$~$\pi^0$ and $0.5$~$\eta$ mesons in
Ref.~\cite{deGouvea:2018cfv}, since additional mesons can be produced by secondary interactions within the $\approx 1$~m long graphite target and other beamline components.


We then simulate the radiative decays $\pi^{0},\eta \rightarrow \gamma \chi \bar{\chi}$ with \texttt{BdNMC}~\cite{deNiverville:2016rqh}. In addition to the scalar DM production supported by \texttt{BdNMC}, we added the option to generate fermions. 
We calculate the rate of the scattering process $\chi + \textrm{Ar} \rightarrow \chi + \textrm{Ar} + A'$ inside the LArTPC as a function of the energy of the DM particle and the path traveled inside the detector~\cite{Kycia:2017ota}.
We compare our signal simulation
 to the calculations of Ref.~\cite{deGouvea:2018cfv} and find
good agreement in the kinematics, e.g., the distribution of the $\ee$ opening angle as a function of the energy of each lepton. The cross section of the process shown in Fig.~\ref{fig:diag}b is calculated using \texttt{GenExLight}~\cite{Kycia:2017ota}. We find an agreement better than $1\%$ when comparing these cross sections to calculations obtained with \texttt{MadGraph}~\cite{Alwall:2014hca}.

We use a ``beam-on" data sample to search for the dark-trident signal where the event triggers coincide with 
the arrival time of neutrinos from the NuMI beam. 
The background is modeled considering three contributions.
Beam-on background events that are triggered by a cosmic ray and not a neutrino interaction are modeled by a ``beam-off" sample collected under identical trigger conditions but when no neutrino beam is present. The ``beam-off" sample is scaled so that its normalization corresponds to the number of beam spills of the beam-on sample.
Neutrino-induced background from the NuMI beam is modeled using a \texttt{GENIE}
Monte Carlo  simulation~\cite{Andreopoulos:2015wxa} embedded in the \texttt{LArSoft} software framework~\cite{Snider:2017wjd}.
The ``in-cryostat $\nu$" sample contains interactions of neutrinos with the argon inside the cryostat, and the ``out-of-cryostat $\nu$" sample describes interactions with the material surrounding the detector.  

We reconstruct neutrino interactions and cosmic rays within the argon with a chain of pattern-recognition algorithms, implemented using the \texttt{Pandora} software development kit~\cite{Marshall:2015rfa,MicroBooNE:2017xvs}.
The algorithms use hits that are formed from the waveforms
read out by the charge collection plane and the two induction planes.
Collections of hits are reconstructed as a track, as expected for a minimum ionizing particle, or a shower, consistent with being an electron or photon conversion. 
\begin{table}[ht]
\centering
\caption{Numbers of events that remain after preselection normalized to POT for the data and the background model.}
\begin{tabular}{lcc}
\hline\hline 
Sample & Run 1 (FHC) & Run 3 (RHC) \\
POT  & $2.2 \times 10^{20}$  & $5.0 \times 10^{20}$  \\
 \hline 
Beam-off & $2410$  &  $4826$ \\
In-cryostat $\nu$ & $1262$ &   $2759$  \\
Out-of-cryostat $\nu$ & $\phantom{0}354$ & $\phantom{0}402$ \\
\hline 
Sum of predictions & $4026$  & $7987$  \\
Beam-on (data)  & $4021$ &  $7684$ \\

\hline\hline
\end{tabular}
\label{tab:presel}
\end{table}

We use the results of the \texttt{Pandora} reconstruction to select events that are consistent with the signal hypothesis.
Dark-trident events are frequently reconstructed as a single shower due to the small opening angle of the $\ee$ pairs and, in a few cases, as two showers arising from a common vertex. Background processes that can mimic such signal topologies are neutral current (NC) interactions 
$\nu+ \textrm{Ar} \rightarrow \nu + \pi^0/\eta + X$, 
where the decays $\pi^{0}/\eta\to\gamma\gamma$ are 
reconstructed as an $\ee$ pair.

Each event is therefore required to have at least one vertex, at least one shower, and no tracks. The efficiency of this preselection for a DM signal lies in the range of $(32$--$40)\%$ for masses in the range $(10$--$400)$~MeV. 
We find good agreement between the number of data events and the sum of the predictions for the background processes after this preselection (Table~\ref{tab:presel}).

We use a convolutional neural network (CNN) for discriminating signal and background based on the 
previous development of such algorithms in MicroBooNE for multiple particle identification (MPID)~\cite{MicroBooNE:2020hho}. Convolutional neural networks (CNNs) are deep learning networks that are 
ideally suited for images reconstructed from LArTPC data~\cite{MicroBooNE:2016dpb, MicroBooNE:2022tdj,  MicroBooNE:2023dci}.
The CNN architecture is based on a model for dense images with adaptations for LArTPCs. Convolution filters of size $3\times 3$ allow scanning of the information contained in showers. The output layer has two neurons that correspond to the probability for signal or background.

We only consider images from the charge collection plane, as it has the best signal to noise ratio~\cite{MicroBooNE:2020hho}. Adding information from the induction planes increases the computing time significantly, but has only minimal impact on the performance of the CNN.
The size of each image in pixels corresponds to $3456$~wires multiplied by $6048$~time ticks. To improve processing time, we 
first compress the time axis by a factor of $6$ and then
crop the images around the interaction vertex producing a region of interest (ROI) of $512 \times 512$ pixels.
After compression, each pixel has a resolution of $\approx 3 \times 3$~mm$^2$.

\begin{figure}[ht]
\includegraphics[width=0.35\textwidth]{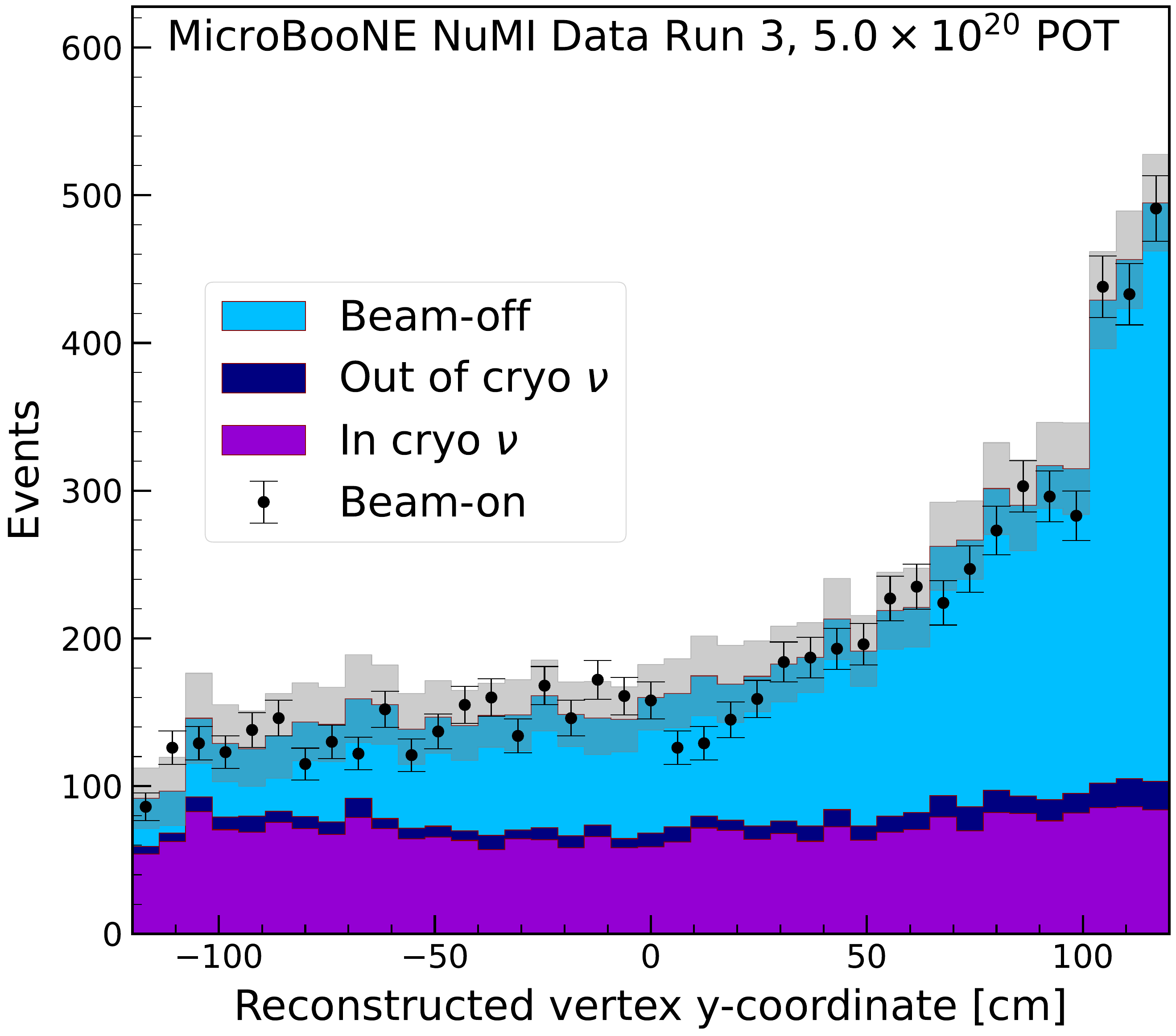}
\caption{\label{fig:yvert} 
Distribution of the $y$ coordinate of the reconstructed vertices for the Run~3 data after preselection 
compared to the background model. The positive direction of the $y$ axis points vertically upwards. The gray band represents the systematic uncertainty in the background model.}
\end{figure}

We validate the agreement of the vertex reconstruction by comparing data and the background model after the preselection (Fig.~\ref{fig:yvert}). The increase of beam-off events towards the top of the detector due to cosmic rays is reproduced by the background model.
While we use the reconstructed vertices for the data and background samples, the true vertex location is used for the training. This prevents the CNN from training on an ROI that does not contain the interaction of interest, which can occur when a vertex is reconstructed at a large distance from the true interaction vertex. 

For the training of the CNN we prepare a dedicated training data set. We use a single signal sample with the parameters $\alpha_{D} = 0.1$, $M_{A^\prime} = 50$~MeV, and $M_{A^\prime}/M_{\chi} = 0.6$. 
As background samples we use cosmic rays simulated with \texttt{CORSIKA}~\cite{Heck:1998vt} and $\nu$ interactions leading to $\pi^{0}$ mesons simulated with \texttt{GENIE}~\cite{Andreopoulos:2015wxa}. 
In addition, we overlay the hits of cosmic rays simulated with \texttt{CORSIKA} 
to the $\nu$ interaction background and the signal samples. Further details on the training of the CNN are provided in the Appendix.

\begin{figure}[htbp]
\includegraphics[width=0.35\textwidth]{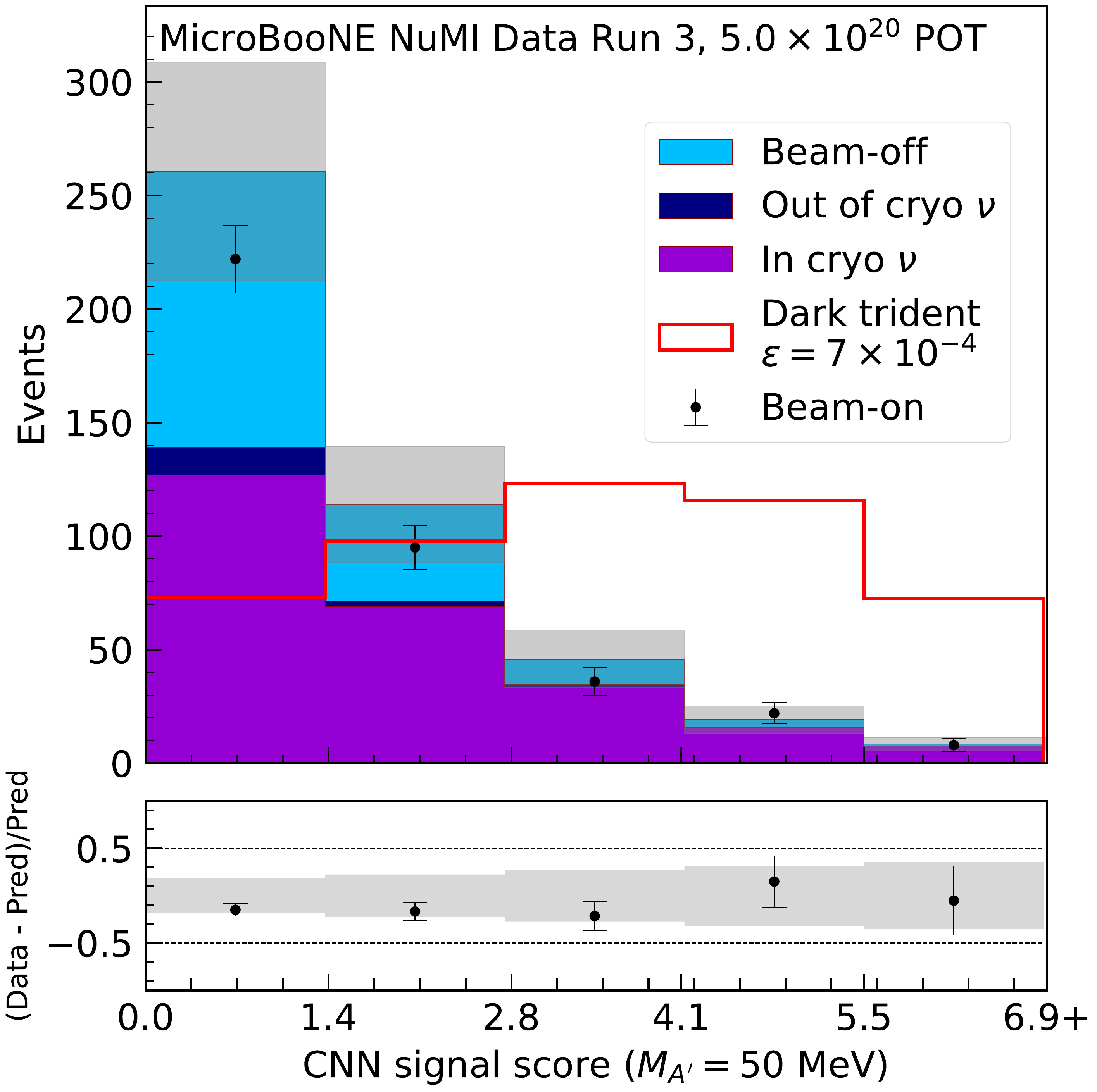}
\caption{\label{fig:score} 
Comparison of the CNN signal score distribution for Run~3 data with the background model after the preselection. The gray band corresponds to the total
systematic uncertainty on the background. The signal distribution for $\alpha_{D} = 0.1$, $M_{A^\prime} = 50$~MeV, and $M_{A^\prime}/M_{\chi} = 0.6$ is superimposed, scaled by an arbitrary factor.}
\end{figure}

\begin{figure}[htbp]
\includegraphics[width=0.35\textwidth]{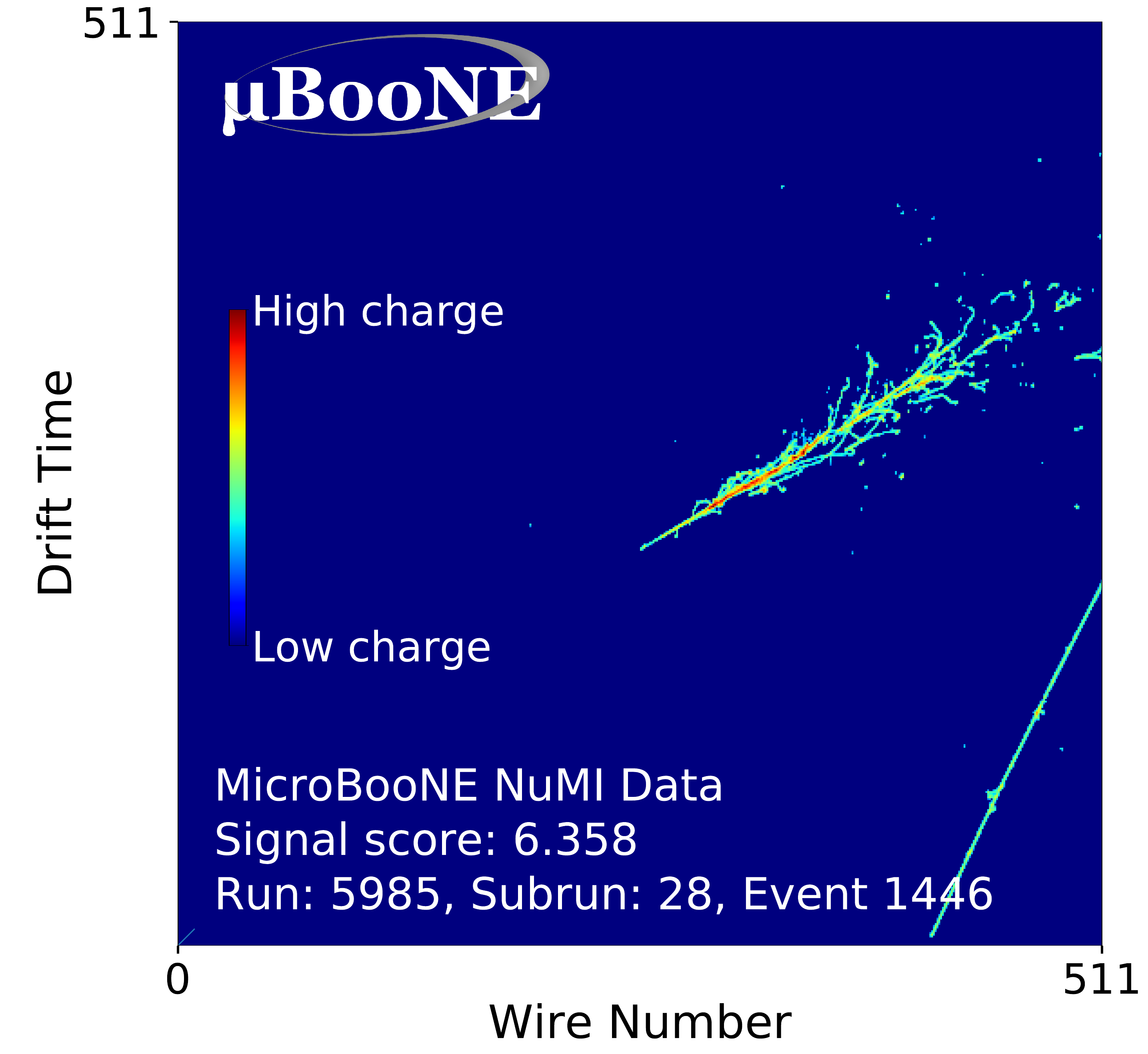}
\caption{\label{fig:display} 
A dark-trident candidate in data with a CNN score of $6.4$, within the ROI of $512\times 512$ pixels ($\approx 1.5 \times 1.5$~m$^2$). A cosmic ray crosses in the lower right-hand corner. }
\end{figure}

The signal MC samples and the samples listed in Tab.~\ref{tab:presel} are passed to the trained CNN model. The events with a CNN signal score $<0$ are rejected obtaining a $99\%$ of background rejection efficiency and signal selection efficiencies in the range $(27$--$30)\%$. Figure~\ref{fig:score} shows the signal score distribution obtained for the events contained in the signal region. A data event with a high CNN signal score is shown in Fig.~\ref{fig:display}, where the shower points in the direction of the NuMI beam. By modifying the training events, we determine that the CNN learns about the orientation of the showers arising from the scattering process.

\begin{figure*}[htbp]
\includegraphics[width=0.4\textwidth]{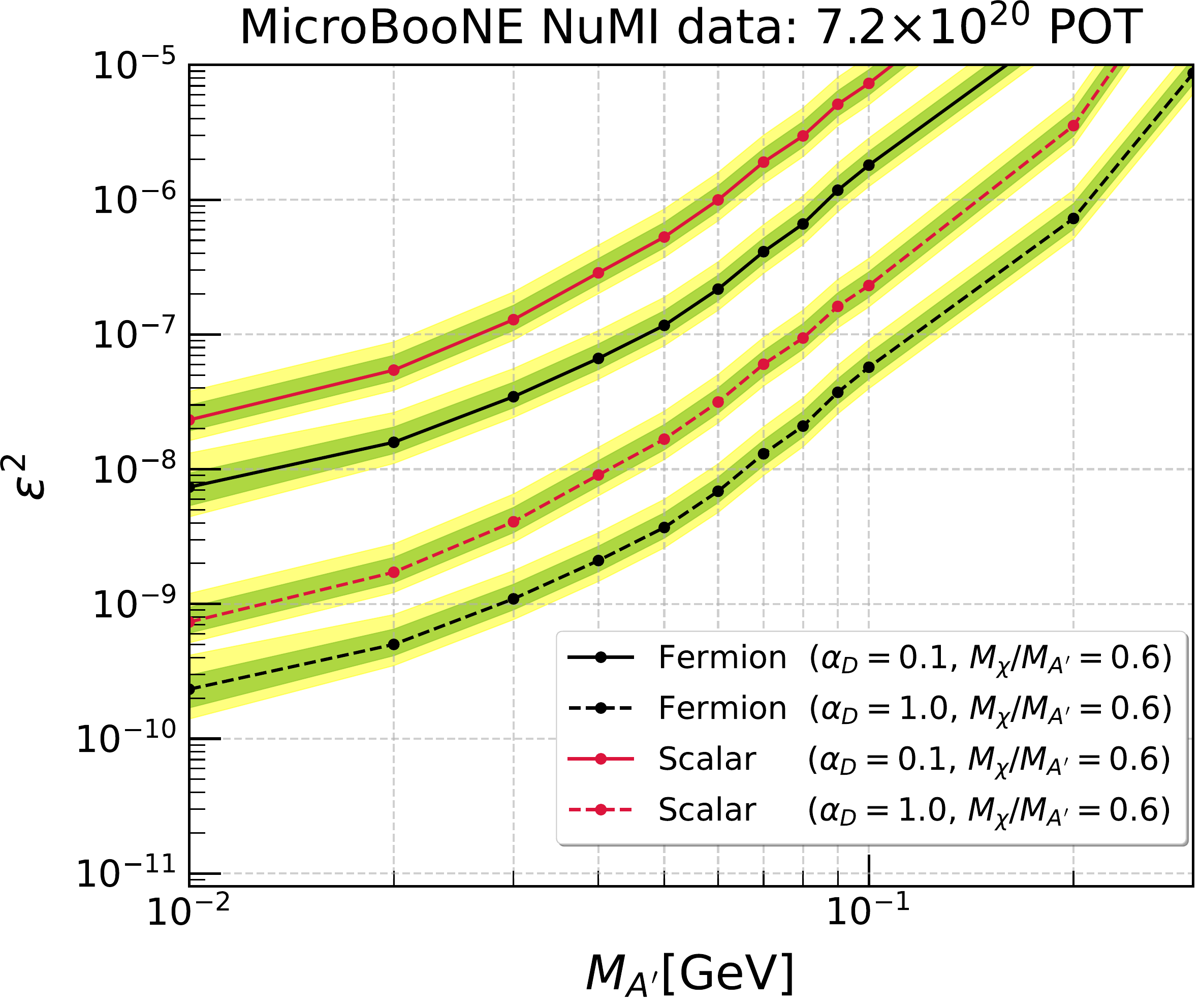}
\includegraphics[width=0.4\textwidth]{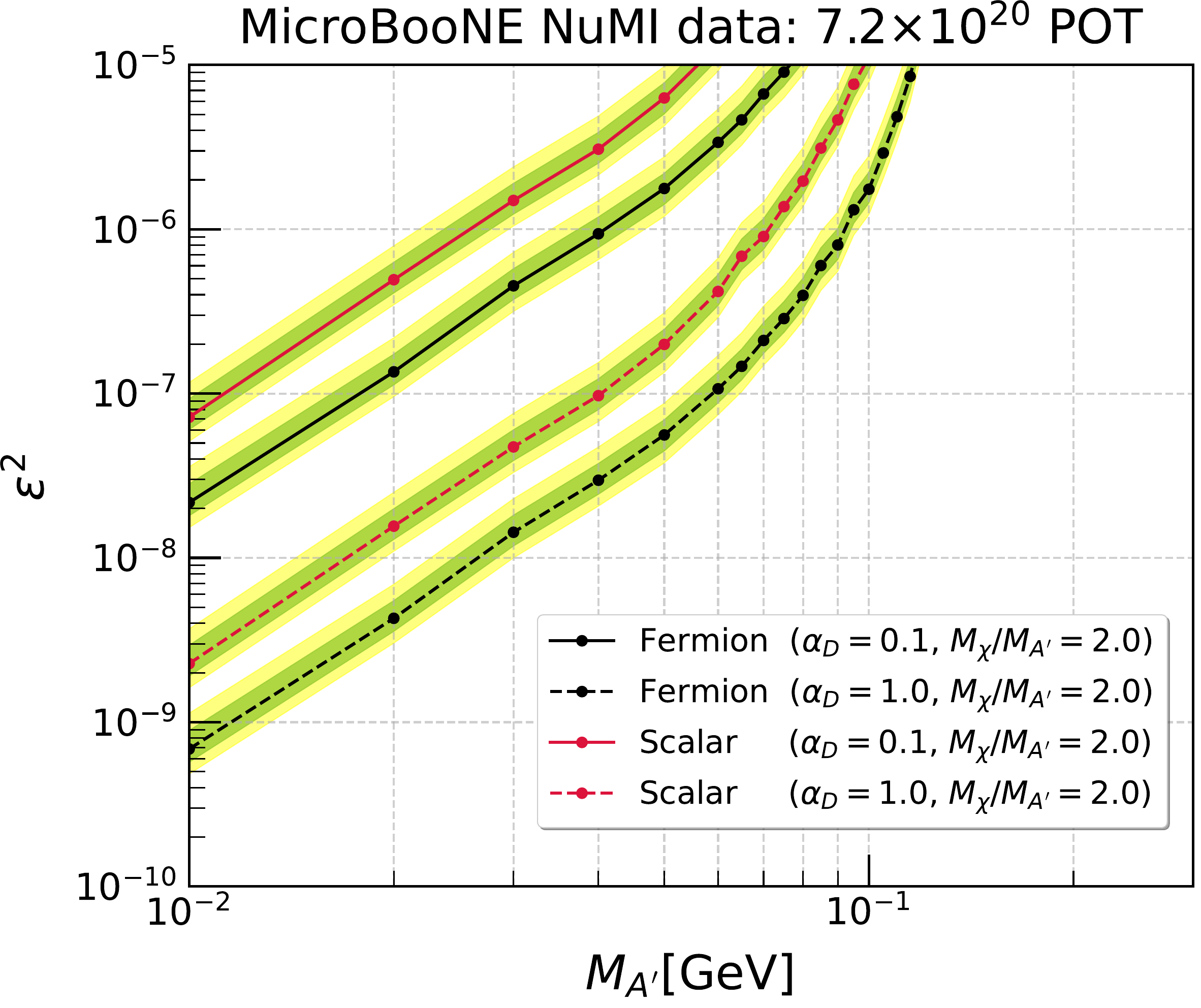} 
\caption{\label{fig:limits} 
The $90\%$~CL observed limits on $\varepsilon^2$ as a function of $\MA$ for $\alpha_D=0.1$ and $\alpha_D=1$, and (a) $\RM=0.6$ and (b) $\RM=2$, together with the $1$ and $2$ standard deviation bands around the median expected limits.
We use a linear interpolation between the mass points.  
A total of $13$ mass values have been simulated for $\RM = 0.6$, equally spaced between $10$--$100$~MeV and between $100$--$400$~MeV and
a total of $19$ mass values for $\RM=2$, an additional $6$ mass values are added at higher $M_A$. 
A table of the limits at each point is provided as supplementary material.}
  \vspace{-77mm}
 \makebox[\textwidth][l]{\hspace{3.2cm}(a) \hspace{6.8cm} (b) }\\ \vspace{77mm}
\end{figure*}

\begin{figure*}[htbp]
\includegraphics[width=0.325\textwidth]
{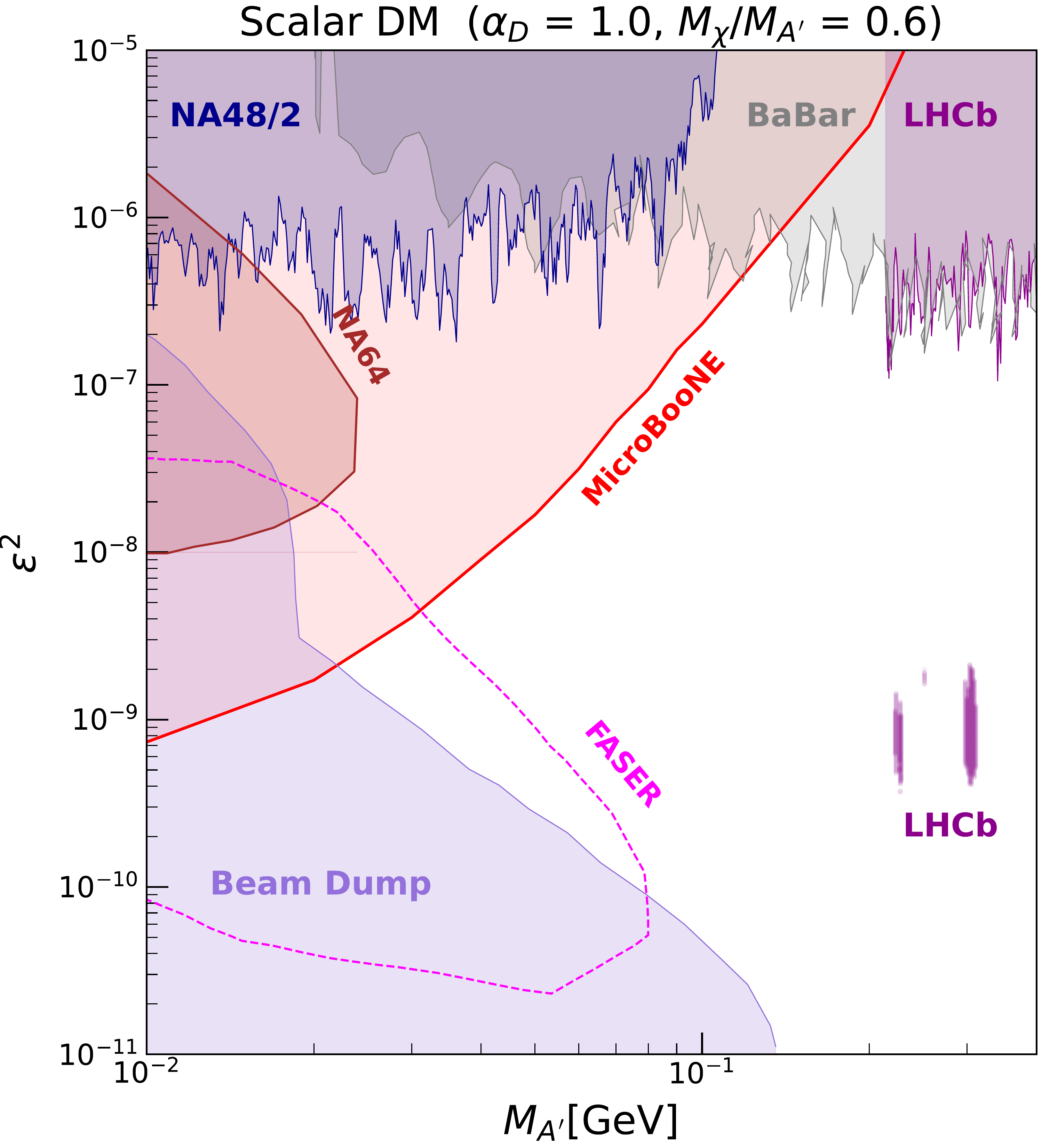}
\includegraphics[width=0.325\textwidth]{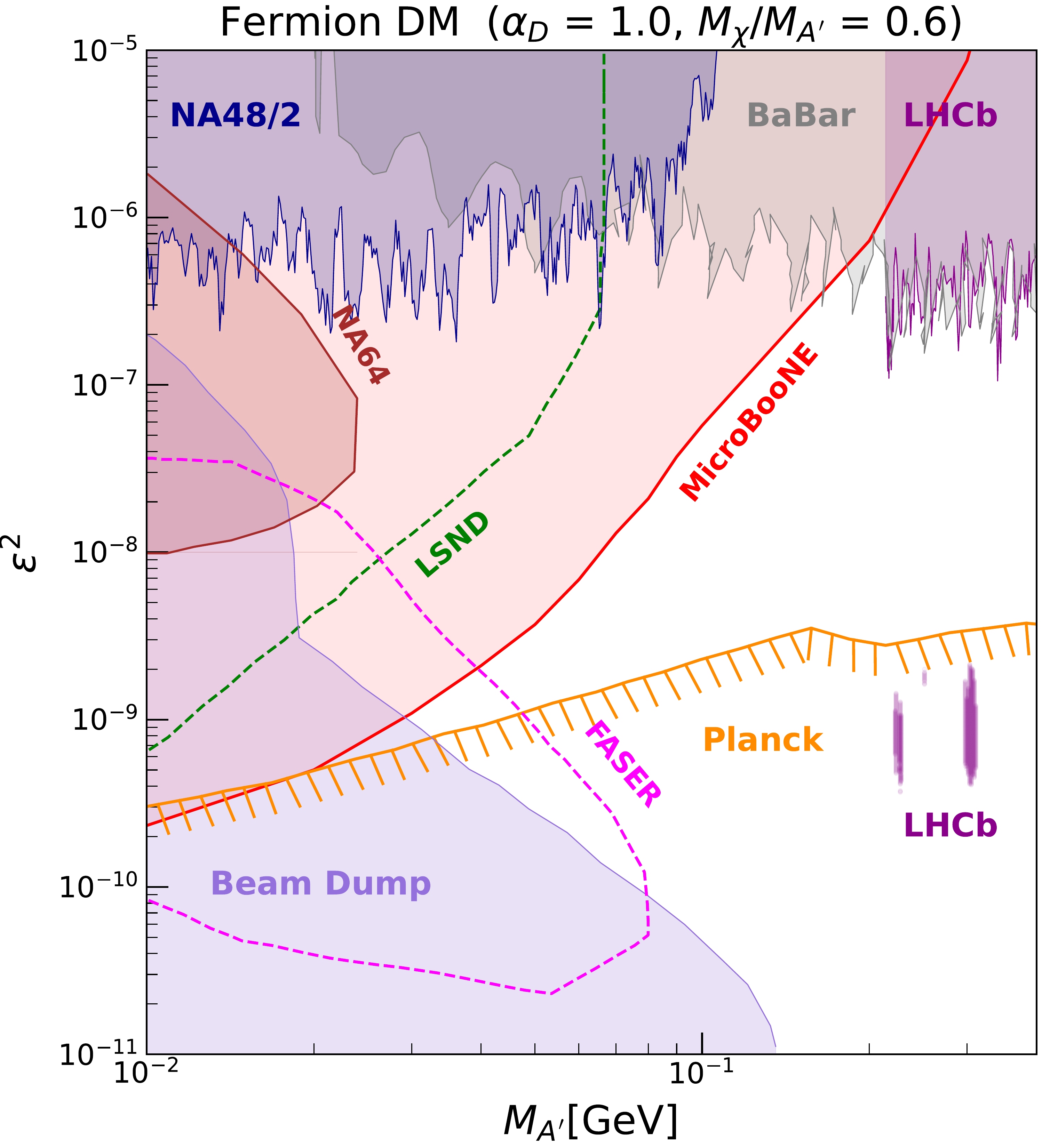}
\includegraphics[width=0.325\textwidth]
{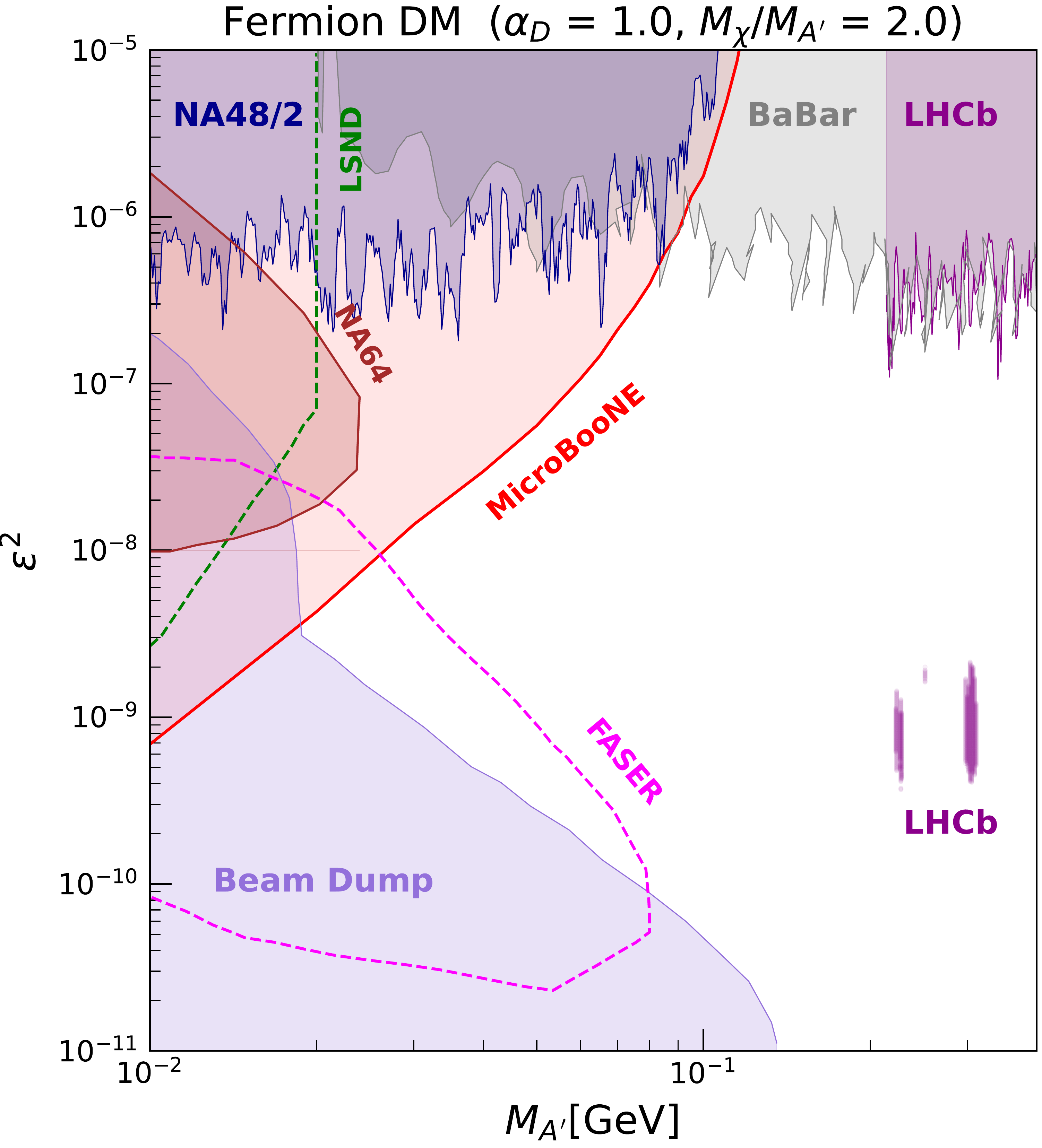}
\caption{\label{fig:results} 
The $90\%$~CL limits on $\varepsilon^2$ as a function of $\MA$ for (a) scalar 
DM with $\alpha_D=1.0$, $\RM=0.6$; (b) fermion DM with $\alpha_D=1.0$, $\RM=0.6$; and (c) for fermion DM with $\alpha_D=1.0$, $\RM=2.0$. The 
constraints provided by the NA48/2~\protect\cite{NA482:2015wmo}, BaBar~\protect\cite{BaBar:2014zli}, 
NA64~\cite{PhysRevD.101.071101}, and LHCb collaborations~\protect\cite{LHCb:2017trq}, and by
beam dump experiments~\cite{Bjorken:1988as,Riordan:1987aw,Bross:1989mp}
are displayed as shaded regions.
The reinterpretations of LSND results~\protect\cite{deGouvea:2018cfv,Kahn:2014sra} and the unpublished 
FASER~\protect\cite{CERN-FASER-CONF-2023-001} limits are shown as dashed lines.
The two isolated contours at $M_{A^\prime} \approx 200$--$300$~MeV are also excluded by LHCb data.
The upper limits on $\varepsilon^2$ from Planck data~\protect\cite{Madhavacheril:2013cna,Slatyer:2015jla} apply for fermion DM with $\RM=0.6$.
}
\vspace{-40mm}
 \makebox[\textwidth][l]{\hspace{5.1cm}(a) \hspace{5.3cm} (b) \hspace{5.3cm} (c)}\\ \vspace{35mm}
\end{figure*}


We evaluate systematic uncertainties for each bin of the CNN score distributions  separately for the different signal models and for background.
For the in-cryostat $\nu$ background, we consider the impact of the neutrino flux simulation $(10$--$20)\%$~\cite{Aliaga:2016oaz} and the neutrino-argon cross-section modeling $(12$--$20)\%$~\cite{MicroBooNE:2021ccs}, hadron interactions with argon ($\approx 1\%$)~\cite{Calcutt_2021}, and detector modeling ($\approx 30\%$)~\cite{MicroBooNE:2021roa}. The beam-off sample is taken from data and therefore has no associated systematic uncertainties other than statistical fluctuations. The impact of the normalization uncertainty on the out-of-cryostat sample and of the POT counting is negligible~\cite{MicroBooNE:2022ctm}.

The sum of the detector-related systematic uncertainties on the signal is in the range $(10$--$20)\%$. 
A form factor accounts for the
spatial distribution of the argon nucleus in the $\chi$-Ar scattering~\cite{deGouvea:2018cfv}. Recalculating the cross sections with different form factors~\cite{DeVries:1987atn, Duda:2006uk} yields uncertainties in the range $(2$--$20)\%$ in the mass range $10\le M_{A'}\le 200$~MeV. 

The signal rate also depends on the NuMI $\pi^0$ and $\eta$ flux
simulated by \texttt{g4NuMI}. We confirm that
the ratio of $\pi^0$ production relative to $\pi^{\pm}$ production in \texttt{g4NuMI} is consistent with expectations of isospin symmetry.
We therefore use the beam flux uncertainty of $22\%$ determined for the charged meson flux~\cite{MicroBooNE:2021gfj}, which includes hadron production and beam line modeling uncertainties.

The CNN score distributions, shown for one model point in Fig.~\ref{fig:score}, are all found to be consistent with the background expectation within uncertainties. We therefore proceed to derive limits on the squared mixing parameter $\varepsilon^2$ as a function of $\MA$. The limit setting is performed with the \texttt{pyhf} algorithm~\cite{pyhf_paper}, which is an implementation of a statistical model to estimate confidence intervals~\cite{Cowan:2010js}. Systematic uncertainties are treated through profile likelihood ratios.
The results are validated with the modified frequentist CL$_s$ calculation of the \texttt{COLLIE} program~\cite{Fisher:2006zz}.
 
The observed limits of Fig.~\ref{fig:limits} are shown at the $90\%$ confidence level (CL$_s=0.1$)
for several benchmark points. Since we use a single CNN model for all signal points, the background CNN score distributions are highly correlated
between the different mass hypotheses $\MA$. All observed limits are therefore consistently
 within the $1$ and $2$ standard deviation ranges around the median expected limit. 

In Fig.~\ref{fig:results}, we compare the results for a scalar dark matter particle $\chi$
with existing constraints on dark-trident processes
from rare pion decays measured by the NA48/2 collaboration~\cite{NA482:2015wmo}, beam dump experiments~\cite{Bjorken:1988as,Riordan:1987aw,Bross:1989mp}, and searches for promptly decaying dark photons into $\ee$ pairs by the BaBar~\cite{BaBar:2014zli}, FASER~\cite{CERN-FASER-CONF-2023-001}, and NA64~\cite{NA64:2021aiq} collaborations. 
The limits obtained by the LHCb collaboration~\cite{LHCb:2017trq}
apply for higher masses $\MA>200$~MeV.
The most sensitive constraints are obtained for $\alpha_D=1$ and $\RM=0.6$. 

For the fermion model, we also compare to reinterpretations of LSND results~\cite{deGouvea:2018cfv,Kahn:2014sra}.
Cosmological constraints on $\chi\bar{\chi}$ annihilation in the early universe are
obtained using Planck measurements on the cosmic microwave background~\cite{Madhavacheril:2013cna,Slatyer:2015jla}. The $\chi\bar{\chi}$ annihilation cross section is only relevant for a fermion $\chi$ and $\RM = 0.6$. The cosmological data 
constrain $\varepsilon^2$ from below, 
since the thermal relic dark-matter density becomes too small for
larger~$\varepsilon^2$~\cite{deGouvea:2018cfv}. 

In summary, we apply convolutional neural networks to obtain first constraints on the production of dark matter in a liquid argon detector exposed to a neutrino beam. The dark matter particles are assumed to interact through the dark photon portal~\cite{deGouvea:2018cfv}.
We consider thermally-produced fermion and boson dark-matter particles $\chi$ with $\RM=0.6$ and $\RM=2$, and
dark fine-structure constants in the range $0.1\le\alpha_D\le 1$. 
The constraints in the plane of the squared kinematic mixing parameter $\varepsilon^2$ and the dark-photon mass $M_{A^\prime}$ exclude previously unexplored regions of parameter space in the range $10\le M_{A^\prime}\le 400$~MeV.

\begin{figure*}[htbp]
    \centering
        \includegraphics[width=0.35\textwidth]{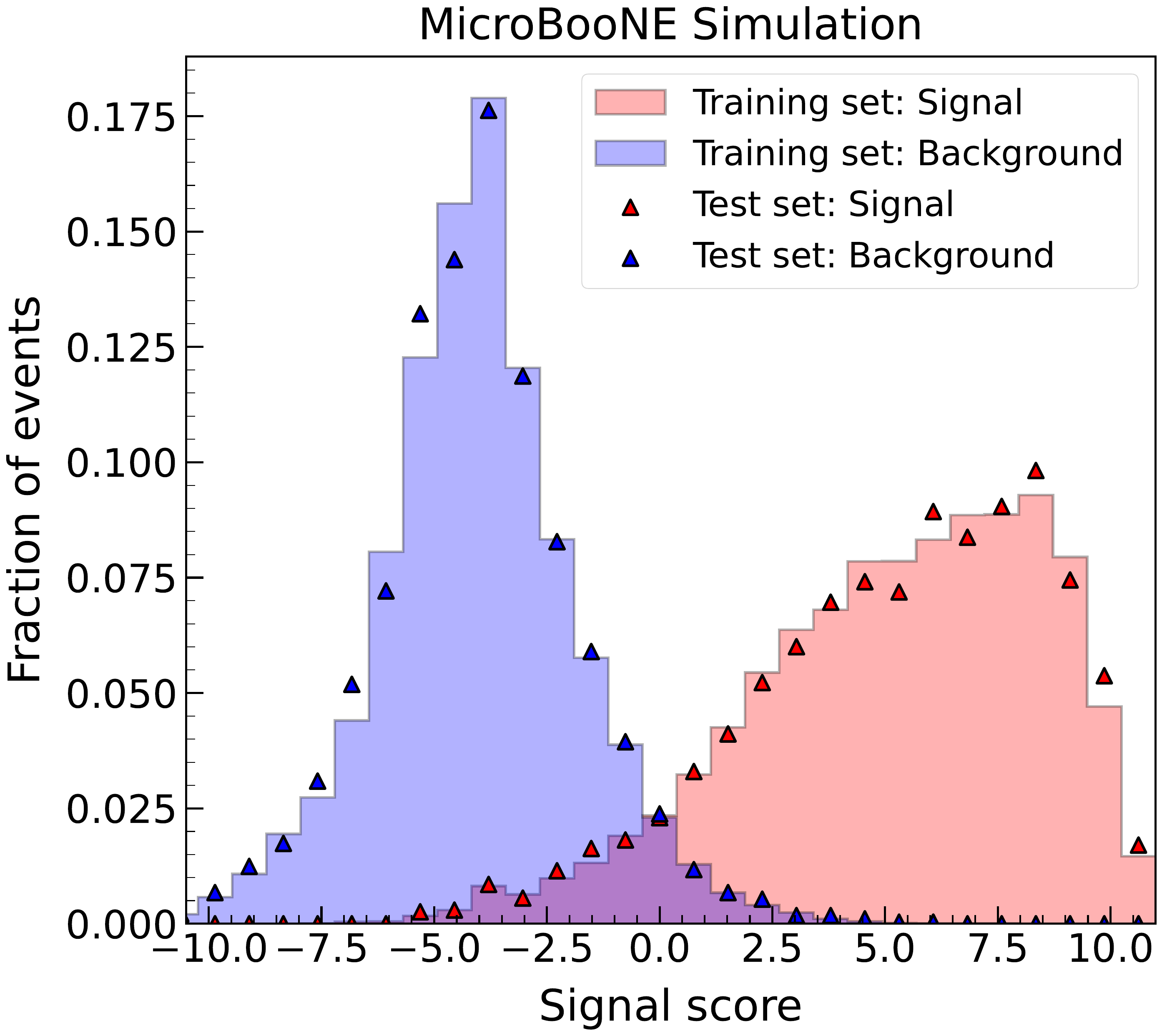} 
        \includegraphics[width=0.34\textwidth]{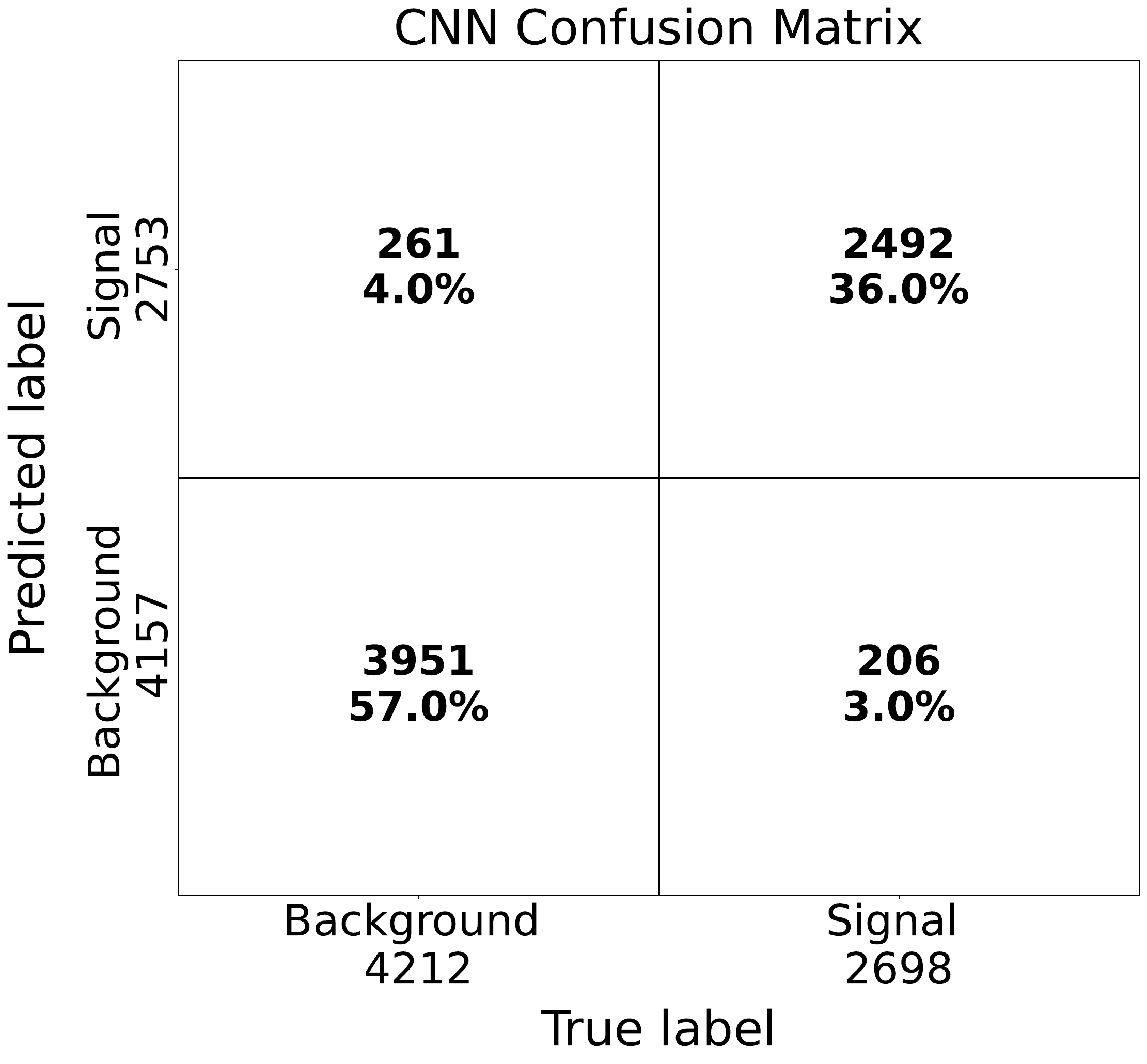} 
           \caption{(a) CNN score distributions for the training and test samples; 
           (b) confusion matrix, obtained for the test set.}
           \label{fig:cnn_training}
          \vspace{-56mm}
 \makebox[\textwidth][l]{\hspace{4.1cm}(a) \hspace{5.4cm} (b)}\\ \vspace{56mm}
\end{figure*}


This document was prepared by the MicroBooNE collaboration using the
resources of the Fermi National Accelerator Laboratory (Fermilab), a
U.S. Department of Energy, Office of Science, HEP User Facility.
Fermilab is managed by Fermi Research Alliance, LLC (FRA), acting
under Contract No. DE-AC02-07CH11359.  MicroBooNE is supported by the
following: 
the U.S. Department of Energy, Office of Science, Offices of High Energy Physics and Nuclear Physics; 
the U.S. National Science Foundation; 
the Swiss National Science Foundation; 
the Science and Technology Facilities Council (STFC), part of the United Kingdom Research and Innovation; 
the Royal Society (United Kingdom); 
the UK Research and Innovation (UKRI) Future Leaders Fellowship; 
and the NSF AI Institute for Artificial Intelligence and Fundamental Interactions. 
Additional support for 
the laser calibration system and cosmic ray tagger were provided by the 
Albert Einstein Center for Fundamental Physics, Bern, Switzerland. We 
also acknowledge the contributions of technical and scientific staff 
to the design, construction, and operation of the MicroBooNE detector 
as well as the contributions of past collaborators to the development 
of MicroBooNE analyses without whom this work would not have been 
possible. 
We also thank the authors of Ref.~\cite{deGouvea:2018cfv} for useful discussions
about the dark-trident model.
For the purpose of open access, the authors have applied 
a Creative Commons Attribution (CC BY) public copyright license to 
any Author Accepted Manuscript version arising from this submission.

\appendix
\section{Appendix}
\label{Appendix}
This Appendix discusses the training and performance of the CNN Classifier in more detail.
The CNN model is trained during $\approx 10{,}000$ iterations ($\approx 5$ epochs) with a batch size of $32$~images and a learning rate of $0.001$~\cite{MicroBooNE:2016dpb}. Dropout layers, regularization terms, and batch normalization are implemented during the training to prevent overfitting. The training progress is monitored with a Binary Cross Entropy (BCE) loss function and using the accuracy, which is defined as the fraction of correctly classified images over the total number of images processed by the CNN. 

\begin{figure}[htbp]
                \includegraphics[width=0.35\textwidth]{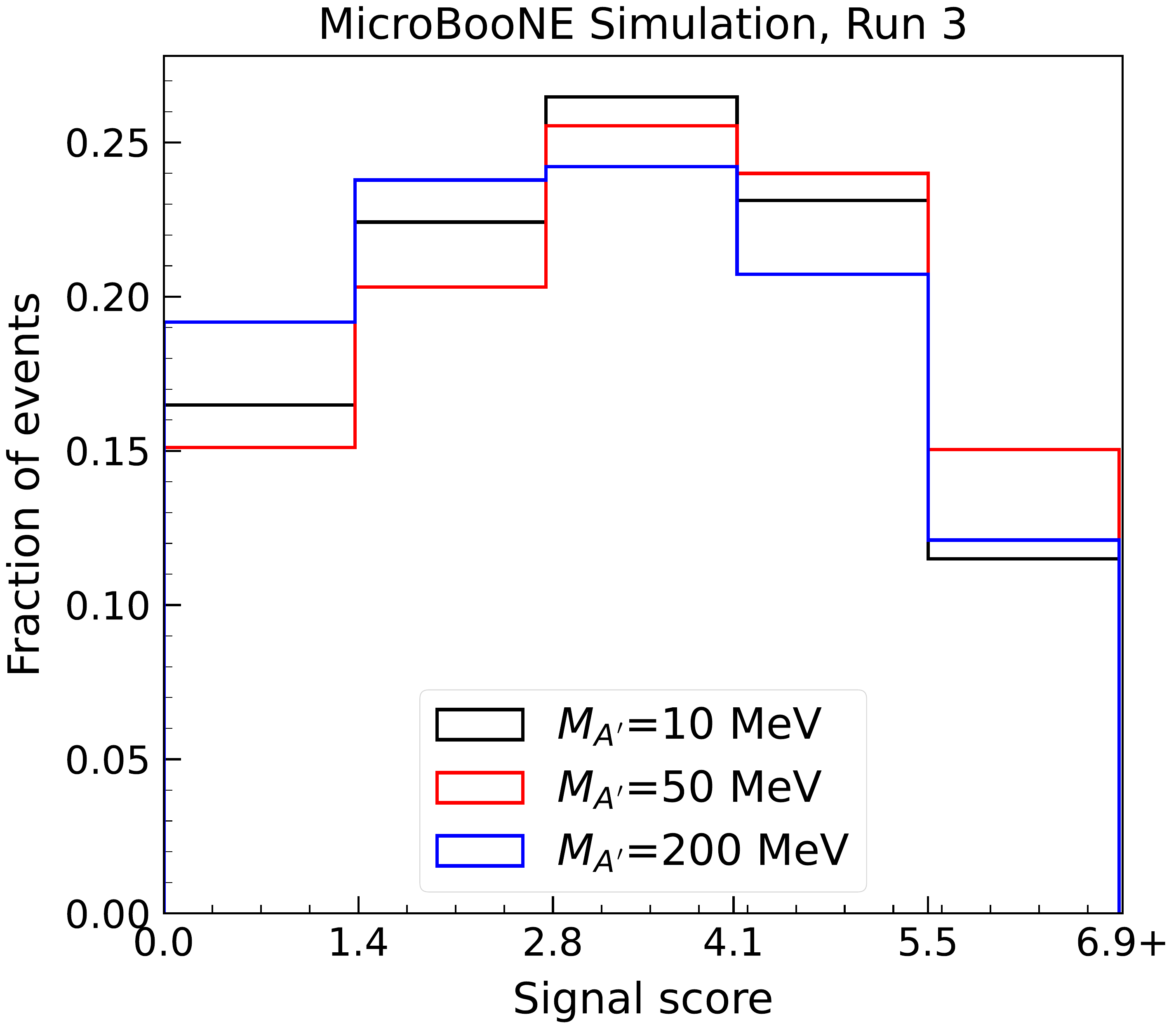} 
           \caption{Signal score distributions for samples simulated with $M_{A^\prime}=10$, $50$, or $200$~MeV.}\label{fig:confusion}
\end{figure}

A test set, comprising $\approx 10\%$ of the events included in the training set, is used to evaluate the progress of the CNN training.  To determine the number of training steps where the CNN model is frozen we use the receiver operating characteristic curve. In Fig.~\ref{fig:cnn_training}a, the distributions of the signal scores obtained for the frozen CNN model are shown for the training and test sets. The distributions show no indication of overfitting. The confusion matrix, obtained after the CNN training for the test set, is given in Fig.~\ref{fig:cnn_training}b. Approximately $93\%$ of events in the test set are labelled correctly. 

The CNN model used in the analysis has been optimized with a benchmark signal point trained against the NC $\pi^0$ and cosmic-ray background samples. Before using the CNN in the analysis, we validate its performance over different signal points and the full background sample (see Table~\ref{tab:presel}). 
The background rejection obtained for the full background sample is not significantly different from the
rejection achieved with the background training sample.
We also find that the shape of the CNN score distribution does not vary much for different signal mass points (Fig.~\ref{fig:confusion}). 


\bibliography{main}

\clearpage
\begin{widetext}
\newenvironment{changemargin}[2]{%
\begin{list}{}{%
\setlength{\topsep}{0pt}%
\setlength{\leftmargin}{#1}%
\setlength{\rightmargin}{#2}%
\setlength{\listparindent}{\parindent}%
\setlength{\itemindent}{\parindent}%
\setlength{\parsep}{\parskip}%
}%
\item[]}{\end{list}}

\section*{Supplemental material}

\vskip 1.0 cm


\subsection*{Additional exclusion contours and tabulated results}

We show exclusion contours for additional model points and give a full list of all results in tabulated form.

\begin{figure*}[htbp]
\includegraphics[width=0.32\textwidth]{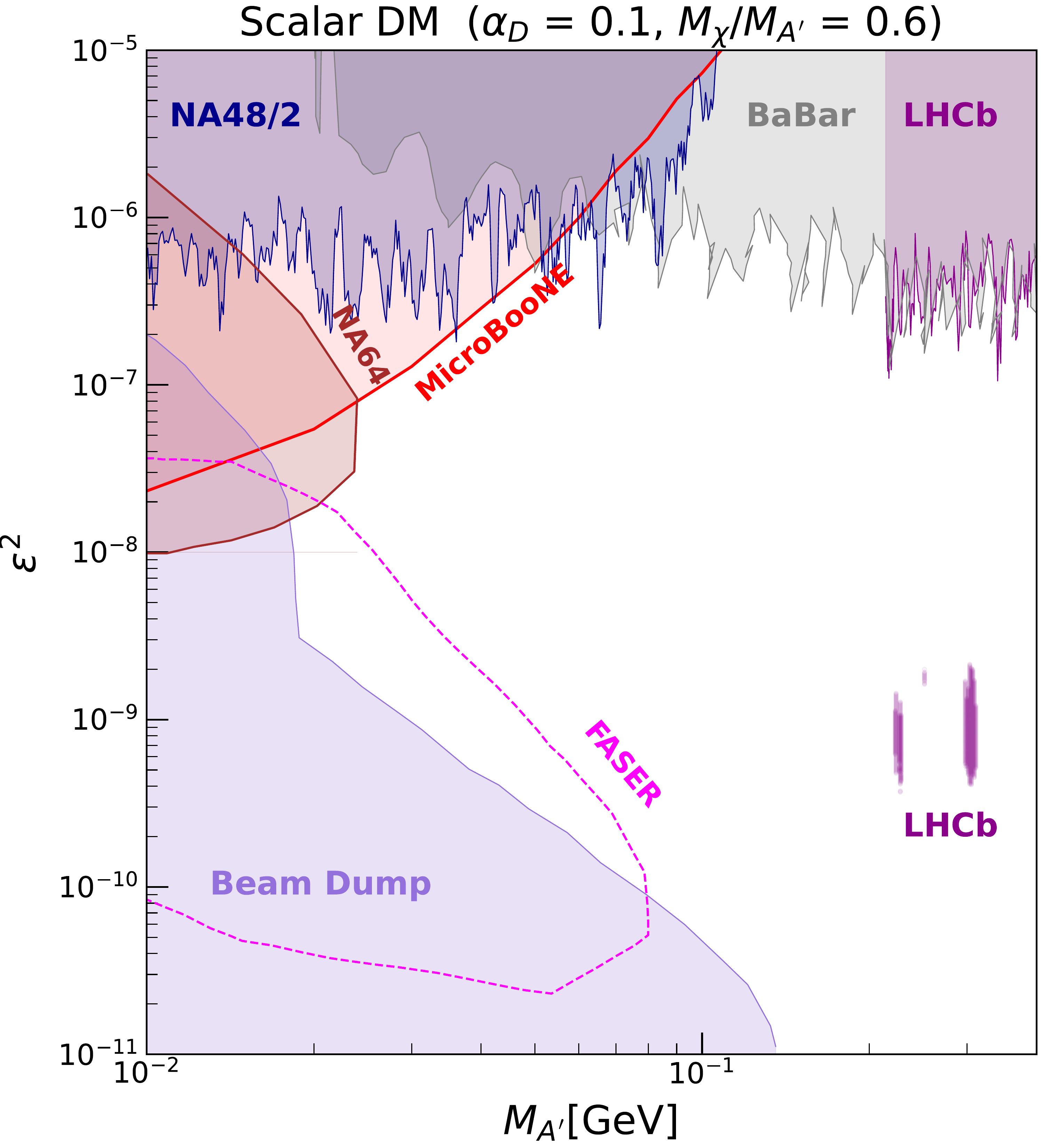}
\includegraphics[width=0.32\textwidth]{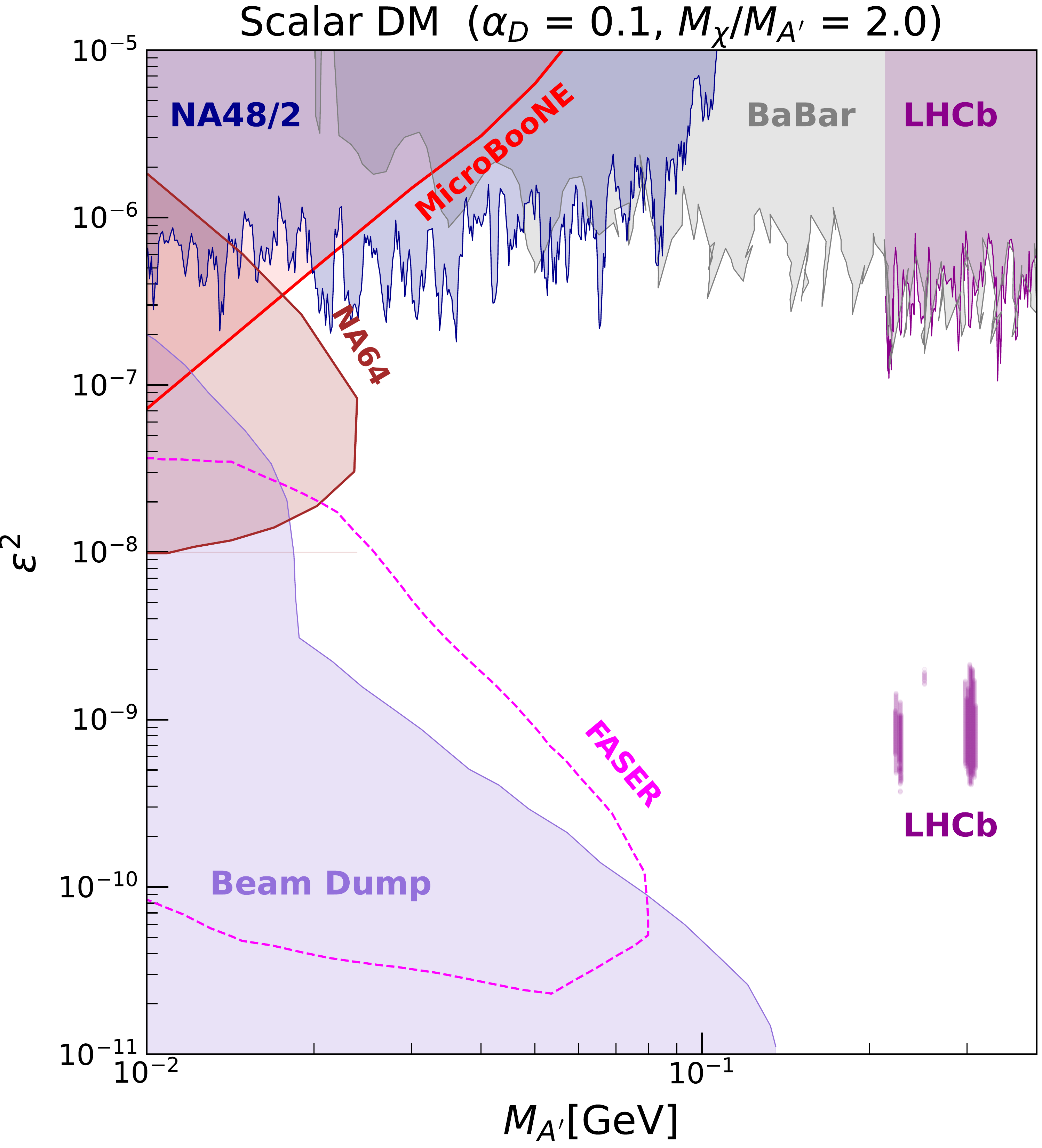}
\includegraphics[width=0.32\textwidth]{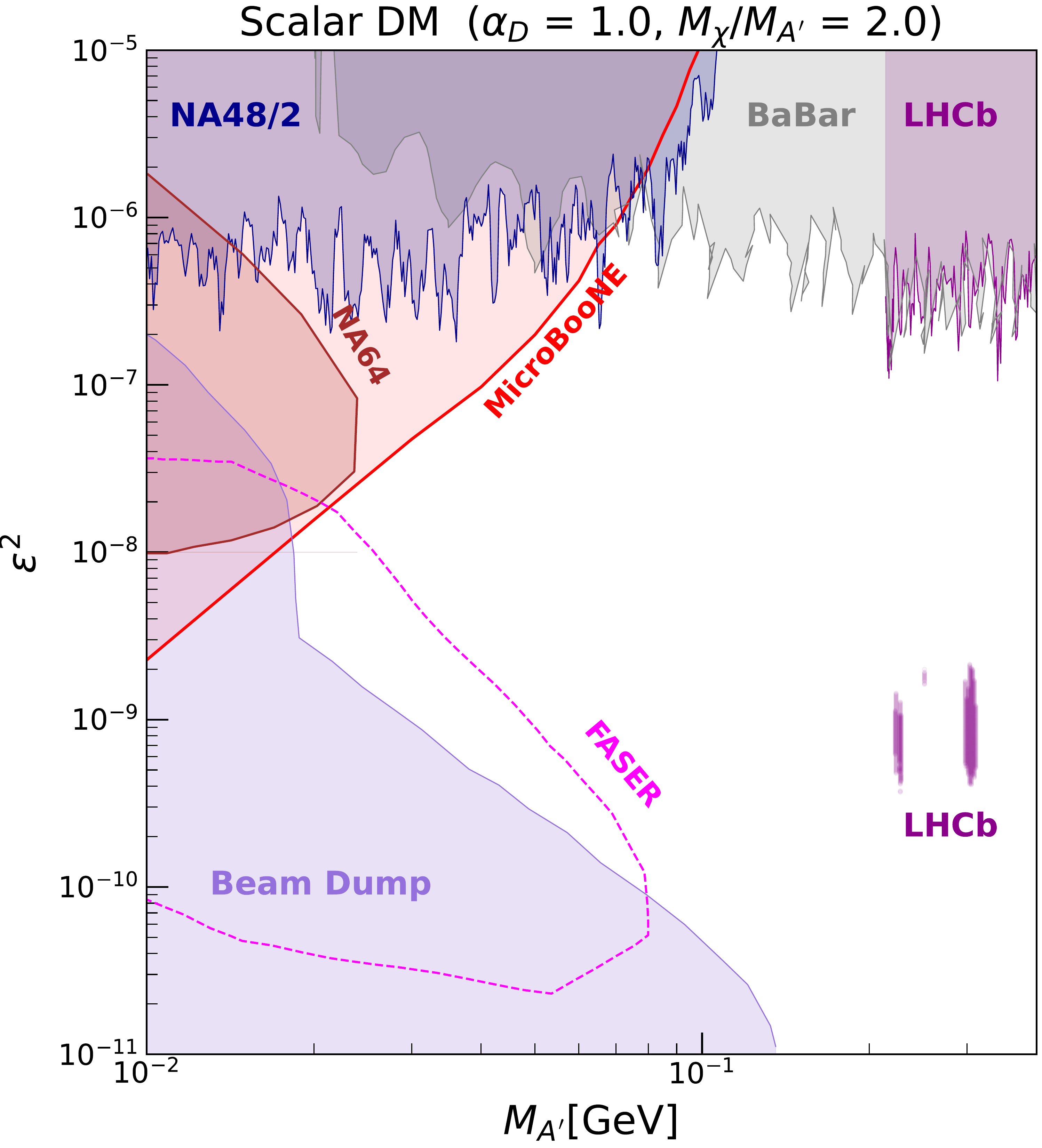}\\
\vskip 0.5cm
\begin{center}
\includegraphics[width=0.32\textwidth]{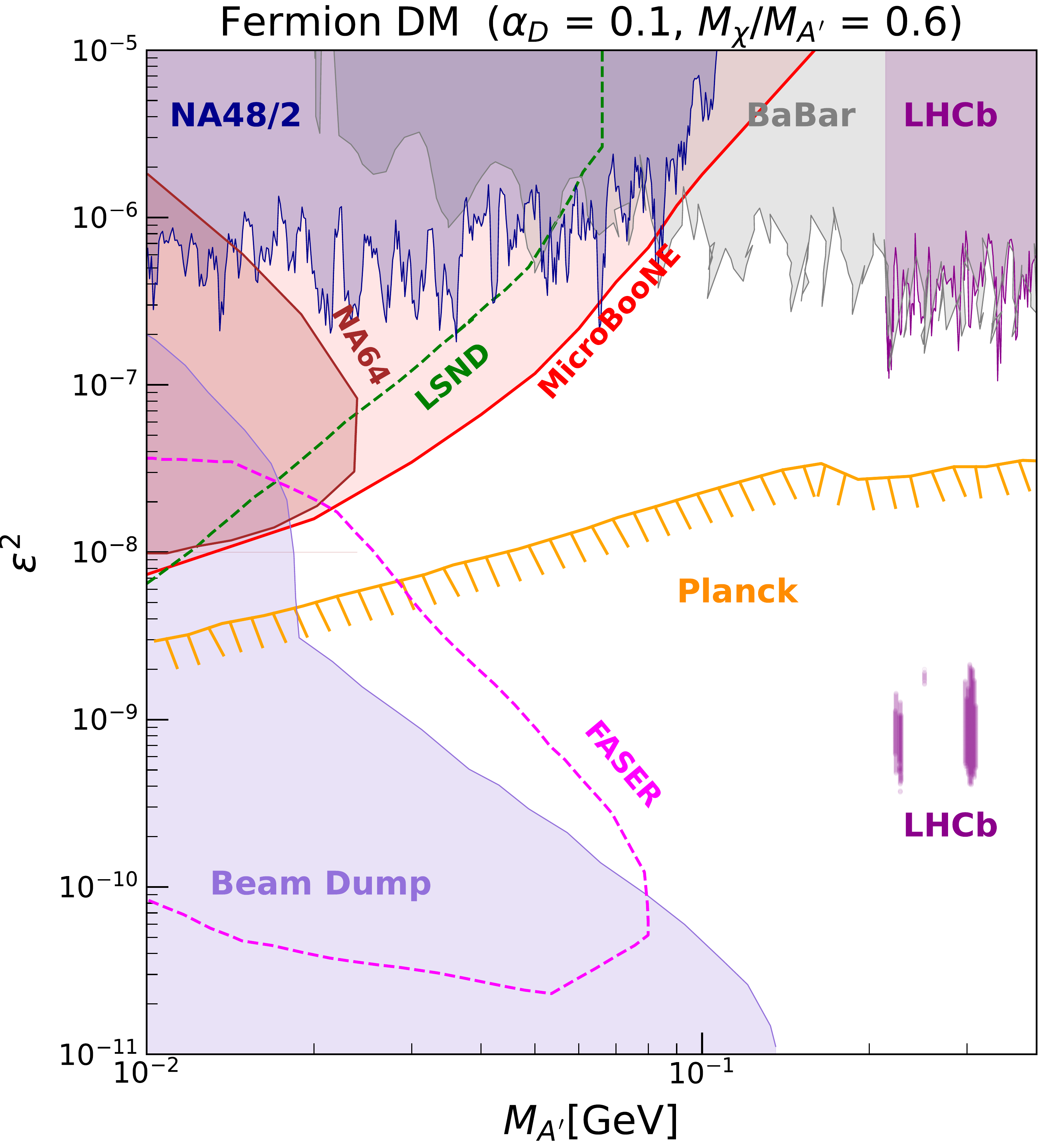}
\includegraphics[width=0.32\textwidth]{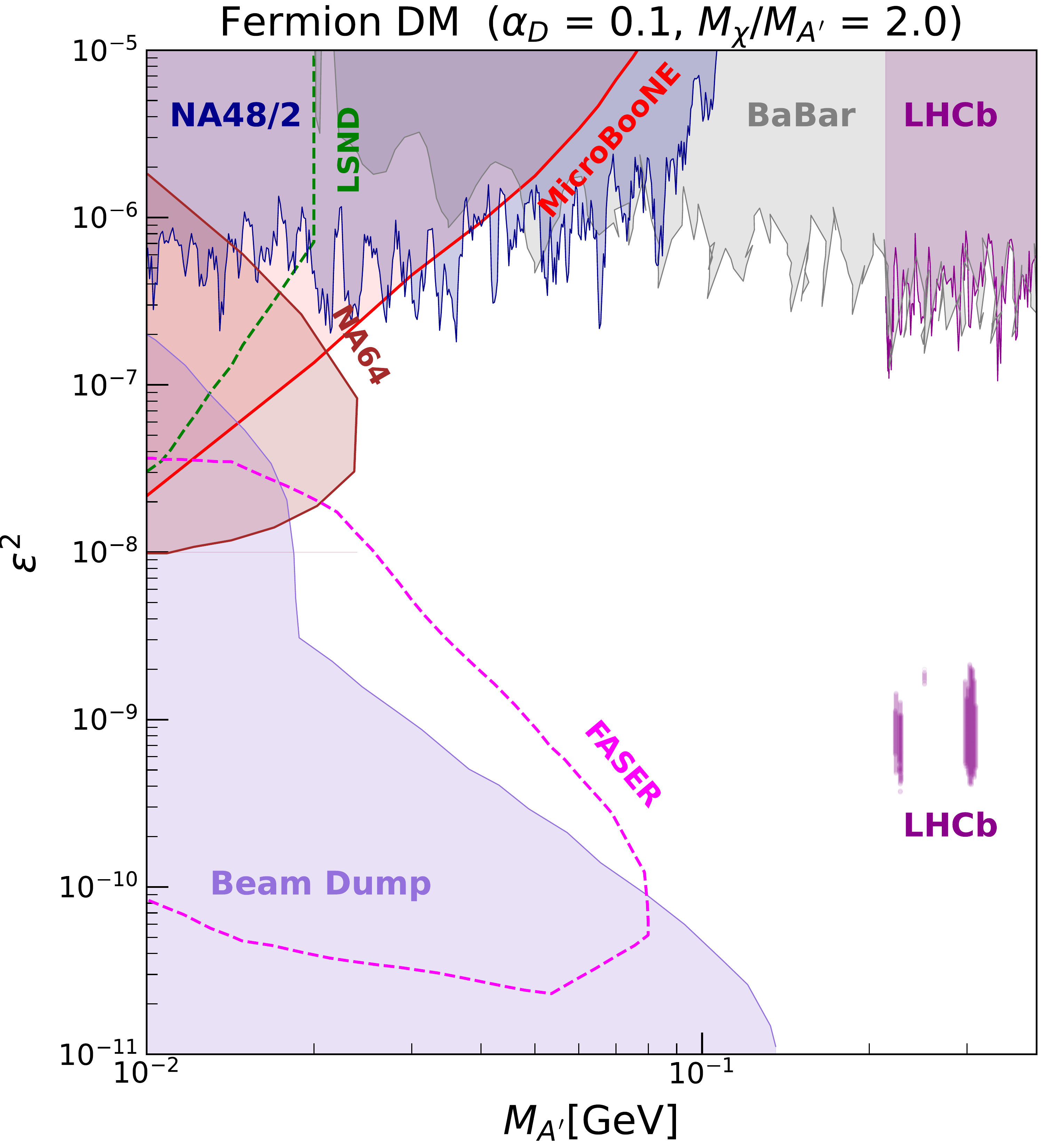}
\end{center}
\caption{\label{fig:limits_fermion} 
The $90\%$~CL limits on $\varepsilon^2$ as a function of $\MA$. The first row corresponds to scalar DM models and the second row to fermion DM models. The 
constraints provided by the NA48/2~\protect\cite{NA482:2015wmo}, BaBar~\protect\cite{BaBar:2014zli}, 
NA64~\cite{NA64:2021aiq}, and LHCb collaborations~\protect\cite{LHCb:2017trq}, and by
beam dump experiments~\cite{Bjorken:1988as,Riordan:1987aw,Bross:1989mp}
are displayed as shaded regions.
The reinterpretations of LSND results~\protect\cite{deGouvea:2018cfv,Kahn:2014sra} and the unpublished 
FASER~\protect\cite{CERN-FASER-CONF-2023-001} limits are shown as dashed lines. 
The two isolated contours at $M_{A^\prime} \approx 200$--$300$~MeV are also excluded by LHCb data.
The upper limits on $\varepsilon^2$ from Planck data~\protect\cite{Madhavacheril:2013cna,Slatyer:2015jla} apply for fermion DM with $\RM=0.6$.}
\end{figure*}

    
\begin{table}[htbp]
\centering
\centerline{Scalar Dark Matter, $\alpha_{D} = 0.1$, $\RM= 2.0$} 
\setlength{\tabcolsep}{7pt} 
\renewcommand{\arraystretch}{1.1} 
\begin{tabular}{c c c c c}
\hline
\hline 
$ M_{A^\prime}$ & Observed & Expected & Expected $1\sigma$ & Expected $2\sigma$ \\ 
(MeV) & & & &  \\
\hline 
10 & 7.19$\cdot 10^{-8}$ & 7.40$\cdot 10^{-8}$ & (6.04--9.29)$\cdot 10^{-8}$ & (0.51--1.17)$\cdot 10^{-7}$  \\ 
20 & 4.93$\cdot 10^{-7}$ & 5.01$\cdot 10^{-7}$ & (4.10--6.30)$\cdot 10^{-7}$ & (3.47--7.92)$\cdot 10^{-7}$  \\ 
30 & 1.50$\cdot 10^{-6}$ & 1.51$\cdot 10^{-6}$ & (1.24--1.90)$\cdot 10^{-6}$ & (1.05--2.39)$\cdot 10^{-6}$  \\ 
40 & 3.07$\cdot 10^{-6}$ & 3.09$\cdot 10^{-6}$ & (2.53--3.89)$\cdot 10^{-6}$ & (2.14--4.89)$\cdot 10^{-6}$  \\ 
50 & 6.30$\cdot 10^{-6}$ & 6.18$\cdot 10^{-6}$ & (5.04--7.78)$\cdot 10^{-6}$ & (4.26--9.81)$\cdot 10^{-6}$  \\ 
60 & 1.32$\cdot 10^{-5}$ & 1.33$\cdot 10^{-5}$ & (1.09--1.68)$\cdot 10^{-5}$ & (0.92--2.11)$\cdot 10^{-5}$  \\ 
65 & 2.17$\cdot 10^{-5}$ & 2.19$\cdot 10^{-5}$ & (1.80--2.76)$\cdot 10^{-5}$ & (1.52--3.47)$\cdot 10^{-5}$  \\ 
70 & 2.86$\cdot 10^{-5}$ & 2.92$\cdot 10^{-5}$ & (2.39--3.67)$\cdot 10^{-5}$ & (2.02--4.62)$\cdot 10^{-5}$  \\ 
75 & 4.35$\cdot 10^{-5}$ & 4.38$\cdot 10^{-5}$ & (3.58--5.51)$\cdot 10^{-5}$ & (3.03--6.93)$\cdot 10^{-5}$  \\ 
80 & 6.21$\cdot 10^{-5}$ & 6.28$\cdot 10^{-5}$ & (5.13--7.90)$\cdot 10^{-5}$ & (4.34--9.95)$\cdot 10^{-5}$  \\ 
85 & 9.87$\cdot 10^{-5}$ & 1.01$\cdot 10^{-4}$ & (0.82--1.26)$\cdot 10^{-4}$ & (0.70--1.59)$\cdot 10^{-4}$  \\ 
90 & 1.46$\cdot 10^{-4}$ & 1.47$\cdot 10^{-4}$ & (1.20--1.84)$\cdot 10^{-4}$ & (1.01--2.32)$\cdot 10^{-4}$  \\ 
95 & 2.41$\cdot 10^{-4}$ & 2.45$\cdot 10^{-4}$ & (2.00--3.08)$\cdot 10^{-4}$ & (1.69--3.88)$\cdot 10^{-4}$  \\ 
100 & 3.55$\cdot 10^{-4}$ & 3.59$\cdot 10^{-4}$ & (2.93--4.51)$\cdot 10^{-4}$ & (2.49--5.67)$\cdot 10^{-4}$  \\ 
105 & 5.74$\cdot 10^{-4}$ & 5.80$\cdot 10^{-4}$ & (4.74--7.29)$\cdot 10^{-4}$ & (4.02--9.18)$\cdot 10^{-4}$  \\ 
110 & 1.19$\cdot 10^{-3}$ & 1.22$\cdot 10^{-3}$ & (1.00--1.53)$\cdot 10^{-3}$ & (0.84--1.93)$\cdot 10^{-3}$  \\ 
115 & 2.18$\cdot 10^{-3}$ & 2.19$\cdot 10^{-3}$ & (1.79--2.76)$\cdot 10^{-3}$ & (1.52--3.48)$\cdot 10^{-3}$  \\ 
120 & 5.63$\cdot 10^{-3}$ & 5.69$\cdot 10^{-3}$ & (4.65--7.15)$\cdot 10^{-3}$ & (3.94--9.00)$\cdot 10^{-3}$  \\ 
125 & 1.74$\cdot 10^{-2}$ & 1.80$\cdot 10^{-2}$ & (1.47--2.27)$\cdot 10^{-2}$ & (1.25--2.86)$\cdot 10^{-2}$  \\ 
\hline 
\hline 
\end{tabular} 
\caption{
The $90\%$~CL observed limits on $\varepsilon^2$ for a scalar DM particle $\chi$ as a function of $\MA$ for $\alpha_D=0.1$ and $\RM=2$, together with the $1$ and $2$ standard deviation bands around the median expected limits. The results are obtained with a data set of $7.2 \times 10^{20}$ POT.}
\end{table}

\begin{table}[htbp]
\centering
\centerline{Scalar Dark Matter, $\alpha_{D} = 1.0$, $\RM = 2.0$} 
\setlength{\tabcolsep}{7pt} 
\renewcommand{\arraystretch}{1.1} 
\begin{tabular}{c c c c c}
\hline
\hline 
$ M_{A^\prime}$ & Observed & Expected & Expected $1\sigma$ & Expected $2\sigma$ \\ 
(MeV) & & & &  \\
\hline 
10 & 2.27$\cdot 10^{-9}$ & 2.34$\cdot 10^{-9}$ & (1.91--2.94)$\cdot 10^{-9}$ & (1.62--3.69)$\cdot 10^{-9}$  \\ 
20 & 1.56$\cdot 10^{-8}$ & 1.59$\cdot 10^{-8}$ & (1.30--1.99)$\cdot 10^{-8}$ & (1.10--2.51)$\cdot 10^{-8}$  \\ 
30 & 4.74$\cdot 10^{-8}$ & 4.79$\cdot 10^{-8}$ & (3.91--6.02)$\cdot 10^{-8}$ & (3.31--7.57)$\cdot 10^{-8}$  \\ 
40 & 9.71$\cdot 10^{-8}$ & 9.77$\cdot 10^{-8}$ & (0.80--1.23)$\cdot 10^{-7}$ & (0.68--1.55)$\cdot 10^{-7}$  \\ 
50 & 1.99$\cdot 10^{-7}$ & 1.95$\cdot 10^{-7}$ & (1.59--2.46)$\cdot 10^{-7}$ & (1.35--3.10)$\cdot 10^{-7}$  \\ 
60 & 4.18$\cdot 10^{-7}$ & 4.21$\cdot 10^{-7}$ & (3.44--5.30)$\cdot 10^{-7}$ & (2.91--6.67)$\cdot 10^{-7}$  \\ 
65 & 6.85$\cdot 10^{-7}$ & 6.94$\cdot 10^{-7}$ & (5.68--8.73)$\cdot 10^{-7}$ & (0.48--1.10)$\cdot 10^{-6}$  \\ 
70 & 9.04$\cdot 10^{-7}$ & 9.24$\cdot 10^{-7}$ & (0.75--1.16)$\cdot 10^{-6}$ & (0.64--1.46)$\cdot 10^{-6}$  \\ 
75 & 1.37$\cdot 10^{-6}$ & 1.38$\cdot 10^{-6}$ & (1.13--1.74)$\cdot 10^{-6}$ & (0.96--2.19)$\cdot 10^{-6}$  \\ 
80 & 1.96$\cdot 10^{-6}$ & 1.99$\cdot 10^{-6}$ & (1.62--2.50)$\cdot 10^{-6}$ & (1.37--3.15)$\cdot 10^{-6}$  \\ 
85 & 3.12$\cdot 10^{-6}$ & 3.18$\cdot 10^{-6}$ & (2.60--4.00)$\cdot 10^{-6}$ & (2.20--5.03)$\cdot 10^{-6}$  \\ 
90 & 4.63$\cdot 10^{-6}$ & 4.63$\cdot 10^{-6}$ & (3.78--5.83)$\cdot 10^{-6}$ & (3.21--7.33)$\cdot 10^{-6}$  \\ 
95 & 7.63$\cdot 10^{-6}$ & 7.74$\cdot 10^{-6}$ & (6.32--9.73)$\cdot 10^{-6}$ & (0.54--1.23)$\cdot 10^{-5}$  \\ 
100 & 1.12$\cdot 10^{-5}$ & 1.14$\cdot 10^{-5}$ & (0.93--1.43)$\cdot 10^{-5}$ & (0.79--1.79)$\cdot 10^{-5}$  \\ 
105 & 1.82$\cdot 10^{-5}$ & 1.84$\cdot 10^{-5}$ & (1.50--2.31)$\cdot 10^{-5}$ & (1.27--2.90)$\cdot 10^{-5}$  \\ 
110 & 3.75$\cdot 10^{-5}$ & 3.85$\cdot 10^{-5}$ & (3.15--4.84)$\cdot 10^{-5}$ & (2.67--6.09)$\cdot 10^{-5}$  \\ 
115 & 6.88$\cdot 10^{-5}$ & 6.94$\cdot 10^{-5}$ & (5.66--8.73)$\cdot 10^{-5}$ & (0.48--1.10)$\cdot 10^{-4}$  \\ 
120 & 1.78$\cdot 10^{-4}$ & 1.80$\cdot 10^{-4}$ & (1.47--2.26)$\cdot 10^{-4}$ & (1.25--2.85)$\cdot 10^{-4}$  \\ 
125 & 5.50$\cdot 10^{-4}$ & 5.71$\cdot 10^{-4}$ & (4.66--7.17)$\cdot 10^{-4}$ & (3.95--9.03)$\cdot 10^{-4}$  \\ 
\hline 
\hline 
\end{tabular}
\caption{
The $90\%$~CL observed limits on $\varepsilon^2$ for a scalar DM particle $\chi$ as a function of $\MA$ for $\alpha_D=1$, and $\RM=2$, together with the $1$ and $2$ standard deviation bands around the median expected limits. The results are obtained with a data set of $7.2 \times 10^{20}$ POT.}
\end{table}


\begin{table}[htbp]
\centering
\centerline{Fermion Dark Matter, $\alpha_{D} = 0.1$, $\RM = 2.0$} 
\setlength{\tabcolsep}{7pt} 
\renewcommand{\arraystretch}{1.1} 
\begin{tabular}{c c c c c}
\hline
\hline 
$ M_{A^\prime}$ & Observed & Expected & Expected $1\sigma$ & Expected $2\sigma$ \\
(MeV) & & & &  \\
\hline 
10 & 2.17$\cdot 10^{-8}$ & 2.22$\cdot 10^{-8}$ & (1.81--2.82)$\cdot 10^{-8}$ & (1.53--3.58)$\cdot 10^{-8}$  \\ 
20 & 1.36$\cdot 10^{-7}$ & 1.38$\cdot 10^{-7}$ & (1.13--1.74)$\cdot 10^{-7}$ & (0.96--2.18)$\cdot 10^{-7}$  \\ 
30 & 4.53$\cdot 10^{-7}$ & 4.58$\cdot 10^{-7}$ & (3.74--5.75)$\cdot 10^{-7}$ & (3.16--7.24)$\cdot 10^{-7}$  \\ 
40 & 9.39$\cdot 10^{-7}$ & 9.45$\cdot 10^{-7}$ & (0.77--1.19)$\cdot 10^{-6}$ & (0.65--1.49)$\cdot 10^{-6}$  \\ 
50 & 1.77$\cdot 10^{-6}$ & 1.74$\cdot 10^{-6}$ & (1.42--2.19)$\cdot 10^{-6}$ & (1.20--2.76)$\cdot 10^{-6}$  \\ 
60 & 3.38$\cdot 10^{-6}$ & 3.40$\cdot 10^{-6}$ & (2.78--4.28)$\cdot 10^{-6}$ & (2.35--5.39)$\cdot 10^{-6}$  \\ 
65 & 4.63$\cdot 10^{-6}$ & 4.69$\cdot 10^{-6}$ & (3.84--5.90)$\cdot 10^{-6}$ & (3.25--7.42)$\cdot 10^{-6}$  \\ 
70 & 6.66$\cdot 10^{-6}$ & 6.81$\cdot 10^{-6}$ & (5.56--8.56)$\cdot 10^{-6}$ & (0.47--1.08)$\cdot 10^{-5}$  \\ 
75 & 9.05$\cdot 10^{-6}$ & 9.12$\cdot 10^{-6}$ & (0.75--1.15)$\cdot 10^{-5}$ & (0.63--1.44)$\cdot 10^{-5}$  \\ 
80 & 1.25$\cdot 10^{-5}$ & 1.27$\cdot 10^{-5}$ & (1.04--1.59)$\cdot 10^{-5}$ & (0.88--2.01)$\cdot 10^{-5}$  \\ 
85 & 1.90$\cdot 10^{-5}$ & 1.94$\cdot 10^{-5}$ & (1.58--2.44)$\cdot 10^{-5}$ & (1.34--3.07)$\cdot 10^{-5}$  \\ 
90 & 2.54$\cdot 10^{-5}$ & 2.54$\cdot 10^{-5}$ & (2.07--3.19)$\cdot 10^{-5}$ & (1.76--4.02)$\cdot 10^{-5}$  \\ 
95 & 4.16$\cdot 10^{-5}$ & 4.21$\cdot 10^{-5}$ & (3.44--5.30)$\cdot 10^{-5}$ & (2.91--6.67)$\cdot 10^{-5}$  \\ 
100 & 5.54$\cdot 10^{-5}$ & 5.61$\cdot 10^{-5}$ & (4.58--7.04)$\cdot 10^{-5}$ & (3.88--8.85)$\cdot 10^{-5}$  \\ 
105 & 9.22$\cdot 10^{-5}$ & 9.32$\cdot 10^{-5}$ & (0.76--1.17)$\cdot 10^{-4}$ & (0.64--1.47)$\cdot 10^{-4}$  \\ 
110 & 1.53$\cdot 10^{-4}$ & 1.57$\cdot 10^{-4}$ & (1.29--1.98)$\cdot 10^{-4}$ & (1.09--2.49)$\cdot 10^{-4}$  \\ 
115 & 2.69$\cdot 10^{-4}$ & 2.71$\cdot 10^{-4}$ & (2.21--3.41)$\cdot 10^{-4}$ & (1.88--4.30)$\cdot 10^{-4}$  \\ 
120 & 5.61$\cdot 10^{-4}$ & 5.67$\cdot 10^{-4}$ & (4.63--7.13)$\cdot 10^{-4}$ & (3.92--8.97)$\cdot 10^{-4}$  \\ 
125 & 1.43$\cdot 10^{-3}$ & 1.49$\cdot 10^{-3}$ & (1.22--1.87)$\cdot 10^{-3}$ & (1.03--2.35)$\cdot 10^{-3}$  \\ 
\hline 
\hline 
\end{tabular} 
\caption{
The $90\%$~CL observed limits on $\varepsilon^2$ for a fermion DM particle $\chi$ as a function of $\MA$ for $\alpha_D=0.1$ and $\RM=2$, together with the $1$ and $2$ standard deviation bands around the median expected limits. The results are obtained with a data set of $7.2 \times 10^{20}$ POT.}
\end{table}

\begin{table}[htbp]
\centering
\centerline{Fermion Dark Matter, $\alpha_{D} = 1.0$, $\RM = 2.0$} 
\setlength{\tabcolsep}{7pt} 
\renewcommand{\arraystretch}{1.1} 
\begin{tabular}{c c c c c}
\hline
\hline 
$ M_{A^\prime}$ & Observed & Expected & Expected $1\sigma$ & Expected $2\sigma$ \\
(MeV) & & & &  \\
\hline 
10 & 6.87$\cdot 10^{-10}$ & 7.03$\cdot 10^{-10}$ & (5.71--8.91)$\cdot 10^{-10}$ & (0.48--1.13)$\cdot 10^{-9}$  \\ 
20 & 4.29$\cdot 10^{-9}$ & 4.37$\cdot 10^{-9}$ & (3.58--5.49)$\cdot 10^{-9}$ & (3.03--6.91)$\cdot 10^{-9}$  \\ 
30 & 1.43$\cdot 10^{-8}$ & 1.45$\cdot 10^{-8}$ & (1.18--1.82)$\cdot 10^{-8}$ & (1.00--2.29)$\cdot 10^{-8}$  \\ 
40 & 2.97$\cdot 10^{-8}$ & 2.99$\cdot 10^{-8}$ & (2.44--3.76)$\cdot 10^{-8}$ & (2.07--4.73)$\cdot 10^{-8}$  \\ 
50 & 5.60$\cdot 10^{-8}$ & 5.49$\cdot 10^{-8}$ & (4.48--6.92)$\cdot 10^{-8}$ & (3.79--8.72)$\cdot 10^{-8}$  \\ 
60 & 1.07$\cdot 10^{-7}$ & 1.08$\cdot 10^{-7}$ & (0.88--1.35)$\cdot 10^{-7}$ & (0.74--1.71)$\cdot 10^{-7}$  \\ 
65 & 1.47$\cdot 10^{-7}$ & 1.48$\cdot 10^{-7}$ & (1.21--1.87)$\cdot 10^{-7}$ & (1.03--2.35)$\cdot 10^{-7}$  \\ 
70 & 2.11$\cdot 10^{-7}$ & 2.15$\cdot 10^{-7}$ & (1.76--2.71)$\cdot 10^{-7}$ & (1.49--3.41)$\cdot 10^{-7}$  \\ 
75 & 2.86$\cdot 10^{-7}$ & 2.88$\cdot 10^{-7}$ & (2.36--3.63)$\cdot 10^{-7}$ & (2.00--4.57)$\cdot 10^{-7}$  \\ 
80 & 3.96$\cdot 10^{-7}$ & 4.01$\cdot 10^{-7}$ & (3.27--5.04)$\cdot 10^{-7}$ & (2.77--6.35)$\cdot 10^{-7}$  \\ 
85 & 6.01$\cdot 10^{-7}$ & 6.13$\cdot 10^{-7}$ & (5.01--7.71)$\cdot 10^{-7}$ & (4.24--9.70)$\cdot 10^{-7}$  \\ 
90 & 8.03$\cdot 10^{-7}$ & 8.03$\cdot 10^{-7}$ & (0.66--1.01)$\cdot 10^{-6}$ & (0.56--1.27)$\cdot 10^{-6}$  \\ 
95 & 1.31$\cdot 10^{-6}$ & 1.33$\cdot 10^{-6}$ & (1.09--1.67)$\cdot 10^{-6}$ & (0.92--2.11)$\cdot 10^{-6}$  \\ 
100 & 1.75$\cdot 10^{-6}$ & 1.77$\cdot 10^{-6}$ & (1.45--2.23)$\cdot 10^{-6}$ & (1.23--2.80)$\cdot 10^{-6}$  \\ 
105 & 2.91$\cdot 10^{-6}$ & 2.95$\cdot 10^{-6}$ & (2.41--3.70)$\cdot 10^{-6}$ & (2.04--4.66)$\cdot 10^{-6}$  \\ 
110 & 4.84$\cdot 10^{-6}$ & 4.97$\cdot 10^{-6}$ & (4.06--6.25)$\cdot 10^{-6}$ & (3.45--7.86)$\cdot 10^{-6}$  \\ 
115 & 8.52$\cdot 10^{-6}$ & 8.58$\cdot 10^{-6}$ & (0.70--1.08)$\cdot 10^{-5}$ & (0.59--1.36)$\cdot 10^{-5}$  \\ 
120 & 1.77$\cdot 10^{-5}$ & 1.79$\cdot 10^{-5}$ & (1.46--2.25)$\cdot 10^{-5}$ & (1.24--2.84)$\cdot 10^{-5}$  \\ 
125 & 4.53$\cdot 10^{-5}$ & 4.70$\cdot 10^{-5}$ & (3.84--5.91)$\cdot 10^{-5}$ & (3.26--7.44)$\cdot 10^{-5}$  \\ 
\hline 
\hline 
\end{tabular}
\caption{
The $90\%$~CL observed limits on $\varepsilon^2$ for a fermion DM particle $\chi$ as a function of $\MA$ for $\alpha_D=1$ and $\RM=2$, together with the $1$ and $2$ standard deviation bands around the median expected limits. The results are obtained with a data set of $7.2 \times 10^{20}$ POT.}
\end{table}


\begin{table}[htbp]
\centering
\centerline{Scalar Dark Matter, $\alpha_{D} = 0.1$, $\RM= 0.6$} 
\setlength{\tabcolsep}{7pt} 
\renewcommand{\arraystretch}{1.1} 
\begin{tabular}{c c c c c}
\hline
\hline 
$ M_{A^\prime}$ & Observed & Expected & Expected $1\sigma$ & Expected $2\sigma$ \\
(MeV) & & & &  \\
\hline 
10 & 2.32$\cdot 10^{-8}$ & 2.36$\cdot 10^{-8}$ & (1.92--2.98)$\cdot 10^{-8}$ & (1.63--3.78)$\cdot 10^{-8}$  \\ 
20 & 5.44$\cdot 10^{-8}$ & 5.56$\cdot 10^{-8}$ & (4.54--6.99)$\cdot 10^{-8}$ & (3.85--8.80)$\cdot 10^{-8}$  \\ 
30 & 1.29$\cdot 10^{-7}$ & 1.31$\cdot 10^{-7}$ & (1.07--1.64)$\cdot 10^{-7}$ & (0.91--2.07)$\cdot 10^{-7}$  \\ 
40 & 2.87$\cdot 10^{-7}$ & 2.91$\cdot 10^{-7}$ & (2.37--3.66)$\cdot 10^{-7}$ & (2.01--4.60)$\cdot 10^{-7}$  \\ 
50 & 5.28$\cdot 10^{-7}$ & 5.39$\cdot 10^{-7}$ & (4.40--6.77)$\cdot 10^{-7}$ & (3.73--8.53)$\cdot 10^{-7}$  \\ 
60 & 9.97$\cdot 10^{-7}$ & 1.01$\cdot 10^{-6}$ & (0.82--1.27)$\cdot 10^{-6}$ & (0.70--1.59)$\cdot 10^{-6}$  \\ 
70 & 1.90$\cdot 10^{-6}$ & 1.90$\cdot 10^{-6}$ & (1.55--2.39)$\cdot 10^{-6}$ & (1.31--3.01)$\cdot 10^{-6}$  \\ 
80 & 2.98$\cdot 10^{-6}$ & 3.02$\cdot 10^{-6}$ & (2.46--3.80)$\cdot 10^{-6}$ & (2.08--4.79)$\cdot 10^{-6}$  \\ 
90 & 5.11$\cdot 10^{-6}$ & 5.11$\cdot 10^{-6}$ & (4.18--6.43)$\cdot 10^{-6}$ & (3.54--8.10)$\cdot 10^{-6}$  \\ 
100 & 7.30$\cdot 10^{-6}$ & 7.33$\cdot 10^{-6}$ & (5.98--9.22)$\cdot 10^{-6}$ & (0.51--1.16)$\cdot 10^{-5}$  \\ 
200 & 1.12$\cdot 10^{-4}$ & 1.15$\cdot 10^{-4}$ & (0.94--1.44)$\cdot 10^{-4}$ & (0.79--1.82)$\cdot 10^{-4}$  \\ 
300 & 2.04$\cdot 10^{-3}$ & 2.08$\cdot 10^{-3}$ & (1.70--2.62)$\cdot 10^{-3}$ & (1.43--3.31)$\cdot 10^{-3}$  \\ 
400 & 9.24$\cdot 10^{-2}$ & 9.19$\cdot 10^{-2}$ & (0.75--1.16)$\cdot 10^{-1}$ & (0.63--1.47)$\cdot 10^{-1}$  \\ 
\hline 
\hline 
\end{tabular}
\caption{
The $90\%$~CL observed limits on $\varepsilon^2$ for a scalar DM particle $\chi$ as a function of $\MA$ for $\alpha_D=0.1$ and $\RM=0.6$, together with the $1$ and $2$ standard deviation bands around the median expected limits. The results are obtained with a data set of $7.2 \times 10^{20}$ POT.}
\end{table}

\begin{table}[htbp]
\centering
\centerline{Scalar Dark Matter, $\alpha_{D} = 1.0$, $\RM= 0.6$} 
\setlength{\tabcolsep}{7pt} 
\renewcommand{\arraystretch}{1.1} 
\begin{tabular}{c c c c c}
\hline
\hline 
$ M_{A^\prime}$ & Observed & Expected & Expected $1\sigma$ & Expected $2\sigma$ \\
(MeV) & & & &  \\
\hline 
10 & 7.34$\cdot 10^{-10}$ & 7.46$\cdot 10^{-10}$ & (6.08--9.43)$\cdot 10^{-10}$ & (0.52--1.19)$\cdot 10^{-9}$  \\ 
20 & 1.72$\cdot 10^{-9}$ & 1.76$\cdot 10^{-9}$ & (1.44--2.21)$\cdot 10^{-9}$ & (1.22--2.78)$\cdot 10^{-9}$  \\ 
30 & 4.07$\cdot 10^{-9}$ & 4.14$\cdot 10^{-9}$ & (3.38--5.20)$\cdot 10^{-9}$ & (2.87--6.54)$\cdot 10^{-9}$  \\ 
40 & 9.07$\cdot 10^{-9}$ & 9.20$\cdot 10^{-9}$ & (0.75--1.16)$\cdot 10^{-8}$ & (0.64--1.45)$\cdot 10^{-8}$  \\ 
50 & 1.67$\cdot 10^{-8}$ & 1.70$\cdot 10^{-8}$ & (1.39--2.14)$\cdot 10^{-8}$ & (1.18--2.70)$\cdot 10^{-8}$  \\ 
60 & 3.15$\cdot 10^{-8}$ & 3.18$\cdot 10^{-8}$ & (2.60--4.00)$\cdot 10^{-8}$ & (2.20--5.04)$\cdot 10^{-8}$  \\ 
70 & 6.01$\cdot 10^{-8}$ & 6.00$\cdot 10^{-8}$ & (4.90--7.55)$\cdot 10^{-8}$ & (4.15--9.52)$\cdot 10^{-8}$  \\ 
80 & 9.41$\cdot 10^{-8}$ & 9.55$\cdot 10^{-8}$ & (0.78--1.20)$\cdot 10^{-7}$ & (0.66--1.51)$\cdot 10^{-7}$  \\ 
90 & 1.62$\cdot 10^{-7}$ & 1.61$\cdot 10^{-7}$ & (1.32--2.03)$\cdot 10^{-7}$ & (1.12--2.56)$\cdot 10^{-7}$  \\ 
100 & 2.31$\cdot 10^{-7}$ & 2.32$\cdot 10^{-7}$ & (1.89--2.91)$\cdot 10^{-7}$ & (1.60--3.67)$\cdot 10^{-7}$  \\ 
200 & 3.55$\cdot 10^{-6}$ & 3.63$\cdot 10^{-6}$ & (2.96--4.56)$\cdot 10^{-6}$ & (2.51--5.75)$\cdot 10^{-6}$  \\ 
300 & 6.46$\cdot 10^{-5}$ & 6.58$\cdot 10^{-5}$ & (5.37--8.30)$\cdot 10^{-5}$ & (0.45--1.05)$\cdot 10^{-4}$  \\ 
400 & 2.92$\cdot 10^{-3}$ & 2.91$\cdot 10^{-3}$ & (2.37--3.67)$\cdot 10^{-3}$ & (2.00--4.66)$\cdot 10^{-3}$  \\ 
\hline 
\hline 
\end{tabular}
\caption{
The $90\%$~CL observed limits on $\varepsilon^2$ for a scalar DM particle $\chi$ as a function of $\MA$ for $\alpha_D=1$ and $\RM=0.6$, together with the $1$ and $2$ standard deviation bands around the median expected limits. The results are obtained with a data set of $7.2 \times 10^{20}$ POT.}
\end{table}


\begin{table}[htbp]
\centering
\centerline{Fermion Dark Matter, $\alpha_{D} = 0.1$, $\RM = 0.6$} 
\setlength{\tabcolsep}{7pt} 
\renewcommand{\arraystretch}{1.1} 
\begin{tabular}{c c c c c}
\hline
\hline 
$ M_{A^\prime}$ & Observed & Expected & Expected $1\sigma$ & Expected $2\sigma$ \\
(MeV) & & & &  \\
\hline 
10 & 7.36$\cdot 10^{-9}$ & 6.86$\cdot 10^{-9}$ & (5.36--9.34)$\cdot 10^{-9}$ & (0.44--1.32)$\cdot 10^{-8}$  \\ 
20 & 1.58$\cdot 10^{-8}$ & 1.61$\cdot 10^{-8}$ & (1.31--2.05)$\cdot 10^{-8}$ & (1.10--2.64)$\cdot 10^{-8}$  \\ 
30 & 3.45$\cdot 10^{-8}$ & 3.51$\cdot 10^{-8}$ & (2.86--4.42)$\cdot 10^{-8}$ & (2.42--5.59)$\cdot 10^{-8}$  \\ 
40 & 6.64$\cdot 10^{-8}$ & 6.74$\cdot 10^{-8}$ & (5.50--8.48)$\cdot 10^{-8}$ & (0.46--1.07)$\cdot 10^{-7}$  \\ 
50 & 1.17$\cdot 10^{-7}$ & 1.19$\cdot 10^{-7}$ & (0.97--1.50)$\cdot 10^{-7}$ & (0.82--1.89)$\cdot 10^{-7}$  \\ 
60 & 2.17$\cdot 10^{-7}$ & 2.19$\cdot 10^{-7}$ & (1.78--2.75)$\cdot 10^{-7}$ & (1.51--3.46)$\cdot 10^{-7}$  \\ 
70 & 4.12$\cdot 10^{-7}$ & 4.12$\cdot 10^{-7}$ & (3.36--5.18)$\cdot 10^{-7}$ & (2.84--6.53)$\cdot 10^{-7}$  \\ 
80 & 6.61$\cdot 10^{-7}$ & 6.72$\cdot 10^{-7}$ & (5.48--8.46)$\cdot 10^{-7}$ & (0.46--1.07)$\cdot 10^{-6}$  \\ 
90 & 1.18$\cdot 10^{-6}$ & 1.18$\cdot 10^{-6}$ & (0.96--1.48)$\cdot 10^{-6}$ & (0.81--1.87)$\cdot 10^{-6}$  \\ 
100 & 1.81$\cdot 10^{-6}$ & 1.81$\cdot 10^{-6}$ & (1.48--2.28)$\cdot 10^{-6}$ & (1.25--2.87)$\cdot 10^{-6}$  \\ 
200 & 2.30$\cdot 10^{-5}$ & 2.35$\cdot 10^{-5}$ & (1.91--2.95)$\cdot 10^{-5}$ & (1.62--3.73)$\cdot 10^{-5}$  \\ 
300 & 2.75$\cdot 10^{-4}$ & 2.80$\cdot 10^{-4}$ & (2.28--3.53)$\cdot 10^{-4}$ & (1.93--4.46)$\cdot 10^{-4}$  \\ 
400 & 8.13$\cdot 10^{-3}$ & 8.09$\cdot 10^{-3}$ & (0.66--1.02)$\cdot 10^{-2}$ & (0.56--1.30)$\cdot 10^{-2}$  \\ 
\hline 
\hline 
\end{tabular}
\caption{
The $90\%$~CL observed limits on $\varepsilon^2$ for a fermion DM particle $\chi$ as a function of $\MA$ for $\alpha_D=1.0$ and $\RM=0.6$, together with the $1$ and $2$ standard deviation bands around the median expected limits. The results are obtained with a data set of $7.2 \times 10^{20}$ POT.}
\end{table}

\begin{table}[htbp]
\centering
\centerline{Fermion Dark Matter, $\alpha_{D} = 1.0$, $\RM = 0.6$} 
\setlength{\tabcolsep}{7pt} 
\renewcommand{\arraystretch}{1.1} 
\begin{tabular}{c c c c c}
\hline
$ M_{A^\prime}$ & Observed & Expected & Expected $1\sigma$ & Expected $2\sigma$ \\
(MeV) & & & &  \\
\hline 
\hline 
10 & 2.33$\cdot 10^{-10}$ & 2.17$\cdot 10^{-10}$ & (1.70--2.95)$\cdot 10^{-10}$ & (1.40--4.16)$\cdot 10^{-10}$  \\ 
20 & 5.01$\cdot 10^{-10}$ & 5.10$\cdot 10^{-10}$ & (4.13--6.49)$\cdot 10^{-10}$ & (3.49--8.33)$\cdot 10^{-10}$  \\ 
30 & 1.09$\cdot 10^{-9}$ & 1.11$\cdot 10^{-9}$ & (0.90--1.40)$\cdot 10^{-9}$ & (0.77--1.77)$\cdot 10^{-9}$  \\ 
40 & 2.10$\cdot 10^{-9}$ & 2.13$\cdot 10^{-9}$ & (1.74--2.68)$\cdot 10^{-9}$ & (1.47--3.38)$\cdot 10^{-9}$  \\ 
50 & 3.69$\cdot 10^{-9}$ & 3.77$\cdot 10^{-9}$ & (3.08--4.74)$\cdot 10^{-9}$ & (2.61--5.97)$\cdot 10^{-9}$  \\ 
60 & 6.86$\cdot 10^{-9}$ & 6.92$\cdot 10^{-9}$ & (5.64--8.70)$\cdot 10^{-9}$ & (0.48--1.09)$\cdot 10^{-8}$  \\ 
70 & 1.30$\cdot 10^{-8}$ & 1.30$\cdot 10^{-8}$ & (1.06--1.64)$\cdot 10^{-8}$ & (0.90--2.06)$\cdot 10^{-8}$  \\ 
80 & 2.09$\cdot 10^{-8}$ & 2.13$\cdot 10^{-8}$ & (1.73--2.67)$\cdot 10^{-8}$ & (1.47--3.37)$\cdot 10^{-8}$  \\ 
90 & 3.72$\cdot 10^{-8}$ & 3.72$\cdot 10^{-8}$ & (3.04--4.68)$\cdot 10^{-8}$ & (2.58--5.90)$\cdot 10^{-8}$  \\ 
100 & 5.71$\cdot 10^{-8}$ & 5.73$\cdot 10^{-8}$ & (4.68--7.21)$\cdot 10^{-8}$ & (3.96--9.08)$\cdot 10^{-8}$  \\ 
200 & 7.26$\cdot 10^{-7}$ & 7.42$\cdot 10^{-7}$ & (6.06--9.34)$\cdot 10^{-7}$ & (0.51--1.18)$\cdot 10^{-6}$  \\ 
300 & 8.69$\cdot 10^{-6}$ & 8.85$\cdot 10^{-6}$ & (0.72--1.12)$\cdot 10^{-5}$ & (0.61--1.41)$\cdot 10^{-5}$  \\ 
400 & 2.57$\cdot 10^{-4}$ & 2.56$\cdot 10^{-4}$ & (2.08--3.23)$\cdot 10^{-4}$ & (1.76--4.10)$\cdot 10^{-4}$  \\ 
\hline 
\hline 
\end{tabular}
\caption{
The $90\%$~CL observed limits on $\varepsilon^2$ for a fermion DM particle $\chi$ as a function of $\MA$ for $\alpha_D=0.1$ and $\RM=0.6$, together with the $1$ and $2$ standard deviation bands around the median expected limits. The results are obtained with a data set of $7.2 \times 10^{20}$ POT.}
\end{table}


\end{widetext}
\end{document}